\documentclass[hidelinks,11pt]{article}
\pdfoutput=1
\usepackage{mathtools}
\usepackage{amssymb}
\usepackage{times}
\usepackage{graphicx}
\usepackage{color}
\usepackage{longtable}
\usepackage{multirow}
\usepackage{capt-of}
\usepackage[natbibapa]{apacite}
\usepackage{rotating}
\usepackage{bbm}
\usepackage{latexsym}
\usepackage{bm}
\usepackage{lineno}
\usepackage[boxed]{algorithm2e}
\usepackage{float}

\setcitestyle{round}

\newcommand{\captionfonts}{\normalsize}

\makeatletter 
\long\def\@makecaption#1#2{%
\vskip\abovecaptionskip
\sbox\@tempboxa{{\captionfonts #1: #2}}%
\ifdim \wd\@tempboxa >\hsize
{\captionfonts #1: #2\par}
\else
\hbox to\hsize{\hfil\box\@tempboxa\hfil}%
\fi
\vskip\belowcaptionskip}
\makeatother 

\begin{document}
\vspace*{0.35in}

\begin{flushleft}
{\Large
\textbf\newline{Detecting Multiple Change Points Using Adaptive Regression Splines with Application to Neural Recordings}
}
\newline
\\
Hazem Toutounji\textsuperscript{1,2,*},
Daniel Durstewitz\textsuperscript{1,3}
\\
\bigskip
\bf{1} Department of Theoretical Neuroscience, Bernstein Center for 
Computational Neuroscience, Central Institute of Mental Health, Medical Faculty
Mannheim, Heidelberg University, Mannheim, Germany
\\
\bf{2} Institute of Neuroinformatics, University of Zurich and ETH Zurich, Zurich, Switzerland
\\
\bf{3} Faculty of Physics and Astronomy, Heidelberg University, Heidelberg, 
Germany
\\
\bigskip
* hazem@ini.ethz.ch

\end{flushleft}


\section*{Abstract}
Time series, as frequently the case in neuroscience, are rarely stationary, but often exhibit abrupt changes due to attractor transitions or bifurcations in the dynamical systems producing them. A plethora of methods for detecting such \emph{change points} in time series statistics have been developed over the years, in addition to test criteria to evaluate their significance. Issues to consider when developing change point analysis methods include computational demands, difficulties arising from either limited amount of data or a large number of covariates, and arriving at statistical tests with sufficient power to detect as many changes as contained in potentially high-dimensional time series. Here, a general method called \emph{Paired Adaptive Regressors for Cumulative Sum} is developed for detecting multiple change points in the mean of multivariate time series. The method's advantages over alternative approaches are demonstrated through a series of simulation experiments. This is followed by a real data application to neural recordings from rat medial prefrontal cortex during learning. Finally, the method's flexibility to incorporate useful features from state-of-the-art change point detection techniques is discussed, along with potential drawbacks and suggestions to remedy them.

\noindent\textbf{Keywords:} change point, cumulative sum, adaptive regression 
splines, nonstationary, bootstrap test, block-permutation, behaviour, spike counts

\section{Introduction} \label{sec:01}

Stationary data are the exception rather than the rule in many areas of science \citep{Aston2012, Elsner2004, Fan2015, Gartner2017, Latimer2015, Paillard1998, Shah2007, Stock2014}. Time series statistics often change, sometimes abruptly, due to transitions in the underlying system dynamics, adaptive processes or external factors. In neuroscience, both behavioural \emph{time series} \citep{Durstewitz2010, Powell2016, Smith2004} and their neural correlates \citep{Durstewitz2010, Gartner2017, Latimer2015, Powell2016, Roitman2002} exhibit strongly nonstationary features which relate to important cognitive processes such as learning \citep{Durstewitz2010, Powell2016, Smith2004} and perceptual decision making \citep{Hanks2017, Latimer2015, Roitman2002}. As such, identifying nonstationary features in behavioural and neural time series becomes necessary, both for interpreting the data in relation to the potential influences generating those features, and for removing those features from the data in order to perform statistical analyses that assume stationary observations \citep{Hamilton1994, Shumway2010}. Abrupt jumps in time series statistics form one important class of nonstationary events. These are often caused by bifurcations, which, in turn, may occur with gradual changes in parameters of the underlying system \citep{Strogatz2001}. Consequently, they are of wide interest to both statistical data analysis and the study of dynamical systems, and are commonly referred to as \emph{change points} \citep[CP;][]{Chen2012}.

Detecting CPs has a long and varied history in statistics, and we will not attempt to exhaustively survey the different approaches, including regression models \citep{Brown1975, Quandt1958}, Bayesian techniques \citep{Chernoff1964} and \emph{cumulative sum} (CUSUM) statistics \citep{Basseville1988, Page1954}, to name but a few, within the limited scope of this article. Instead, we refer the reader to the excellent reviews on the topic \citep{Aminikhanghahi2017, Bhattacharya1994, Chen2012} and focus on the \emph{offline} CUSUM class of methods \citep{Hinkley1971a} to which PARCS belongs \citep[as opposed to sequential CUSUM methods,][that locate a CP online, while the time series is evolving]{Page1954}, specifically methods that aim at detecting CPs in the \emph{mean} of the time series. CUSUM-based methods are powerful, easy to implement, and are backed up by an extensive literature, theoretical results and various extensions to multiple CPs and multivariate scenarios, making them an ideal starting point. These methods assume that the time series is piecewise stationary in the statistic under consideration (e.g., piecewise constant mean) and rely on a cumulative sum transformation of the time series. Commonly, \emph{at-most-one-change} (AMOC) is identified by maximum-type statistics \citep{Kirch2007} at the extremum of the curve resulting from that transformation \citep{Antoch1995, Basseville1988}.

Extending the CUSUM method to multiple CPs usually involves repetitive partitioning of the time series upon each detection \citep[\emph{binary segmentation} methods;][]{Bai1997, Cho2015, Fryzlewicz2014, Olshen2004, Scott1974}. This segmentation procedure, however, may hamper detection in later iterations as the reduction in number of observations depletes statistical power exponentially fast as more CPs are to be retrieved. In this article, we develop the PARCS (\emph{Paired Adaptive Regressors for Cumulative Sum}) method which offers a straightforward extension that leverages the full time series in order to detect multiple CPs, thus providing a new solution to this issue. PARCS rests on the fact that a CUSUM transformation of the data relates to computing an integral transformation of the piecewise constant mean time series model, resulting in a \emph{piecewise linear} mean function that bends at potential CPs and could be approximated by \emph{adaptive regression spline} methods \citep{Friedman1991, Friedman1989, Stone1997}. Namely, rather than attempting to approximate the discontinuous time series mean directly \citep{Efron2004, Vert2010}, the PARCS model is an approximation to the continuous CUSUM-transformed time series by a piecewise linear function. The bending points of the PARCS model are each defined by a pair of non-overlapping piecewise linear regression splines that are first selected by a two-stage iterative procedure.

The PARCS model is further refined by a nonparametric CP significance test based on bootstraps \citep{Antoch2001, Dumbgen1991, Huskova2004, Kirch2007, Matteson2014}. While analytically derived parametric tests may usually be preferable over bootstrap-based tests due to better convergence and coverage of the tails, in the current CP setting closed form expressions for parametric tests are hard to come by and are usually replaced by approximations \citep{Gombay1996, Horvath1997}. In this case, tests based on bootstraps are preferable since they are known to converge faster to the limit distribution of the test statistic \citep[often they are also not as conservative as parametric approximations for datasets of a relatively small size;][]{Antoch2001, Csorgo1997, Kirch2007}. In order to accommodate the possibility of \emph{temporally dependent} noise in the data \citep{Antoch1997, Horvath1997, Picard1985}, model selection is carried out by a nonparametric \emph{block-permutation} bootstrap procedure \citep{Davison1997, Huvskova2001, Kirch2007} developed specifically for PARCS, which relies on a test statistic that quantifies the amount of bending at each candidate CP. Since model estimation is based on linear regression, PARCS is also effortlessly extended to spatially independent, \emph{multivariate} time series.

The article is structured as follows. Section \ref{sec:0201} introduces the CUSUM method for AMOC detection. We then develop the PARCS method, presenting in Section \ref{sec:0202} the procedure for inferring a nested model that allows for significance testing of multiple CPs, followed in Section \ref{sec:0203} by an outline of the nonparametric permutation test procedure for refining the PARCS model further. Results in Section \ref{sec:03} illustrate that PARCS improves on several issues inherent in classical methods for change point analysis. In Section \ref{sec:0301}, we compare the PARCS approach to the CUSUM method in detecting a single CP, followed in Section \ref{sec:0302} by a comparison with standard binary segmentation in detecting multiple CPs. We also demonstrate in Section\ref{sec:0303} that PARCS is successful in detecting CPs in spatially independent, multivariate time series. We then present in Section \ref{sec:0304} an example from the neurosciences, in which neural and behavioural CPs are compared during operant rule-switching learning \citep{Durstewitz2010}. Finally, we discuss in Section \ref{sec:04} the PARCS approach in relation to other state-of-the-art CP detection methods, along with drawbacks and potential extensions.

\section{Methods} \label{sec:02}

This section outlines the CUSUM method and the PARCS extension to multiple CPs, in addition to a nonparametric permutation technique to test for the statistical significance of CPs as identified by PARCS. For generality, the formulation assumes temporally dependent observations in the time series, independent observations being a special case.

\subsection{CUSUM: Cumulative Sum of Differences to the Mean} \label{sec:0201}

A class of methods for identifying a single CP in the mean relies on computing a CUSUM transformation of the time series $\mathbf{x} = \{x_t\}_{1:T}$. A useful formulation that allows for dependent observations in the time series is given by the moving average (MA) step model \citep{Antoch1997, Horvath1997, Lombard1994, Kirch2007}, 
\begin{equation} \label{eq:model1}
x_t = b + w \cdot \mathbf{1}_{t-c} +
\sum_{\tau \ge 0} \kappa_{\tau} \epsilon_{t-\tau}
\quad ; \quad \kappa_0 = 1 , \;
\epsilon_t \sim \mathcal{N} (0,\sigma^2) ,
\end{equation}
where a jump in the time series mean from \emph{baseline} $b$ to $b+w$ occurs after time step $c$, the change point. The \emph{step parameter} or \emph{weight} $w$ is positive (negative) when the time series mean increases (decreases) following $c$. The largest integer $\tau$ such that noise coefficient $\kappa_\tau \neq 0$ defines a finite order $q$ of the MA process, which is 0 for temporally independent observations. We will assume that the MA process is stationary, which will always be the case if it is finite, with $\epsilon_t$ independent and identically distributed (i.i.d.) random variables \citep[for an infinite process, points $x_t$ for $t \le 0$ may be considered unobserved, and coefficients $\kappa_\tau$ have to fulfil certain conditions to make the process stationary, as given, for instance, in][]{Shumway2010}. The Gaussian noise assumption in the MA process can be relaxed, as long as the noise process has zero mean and finite, constant variance \citep[see][for theoretical results on the more general form of dependent noise]{Antoch1997, Horvath1997, Lombard1994, Kirch2007}. The discrete Heaviside step function, $\mathbf{1}_{t-c}$, is defined by,
\begin{equation*}
\mathbf{1}_{i} =
\begin{cases}
1 \quad & \mathrm{if} \ i > 0 , \\
0 \quad & \mathrm{otherwise} .
\end{cases}
\end{equation*}

\begin{figure}[!t]
\centering
\includegraphics[width=\linewidth]{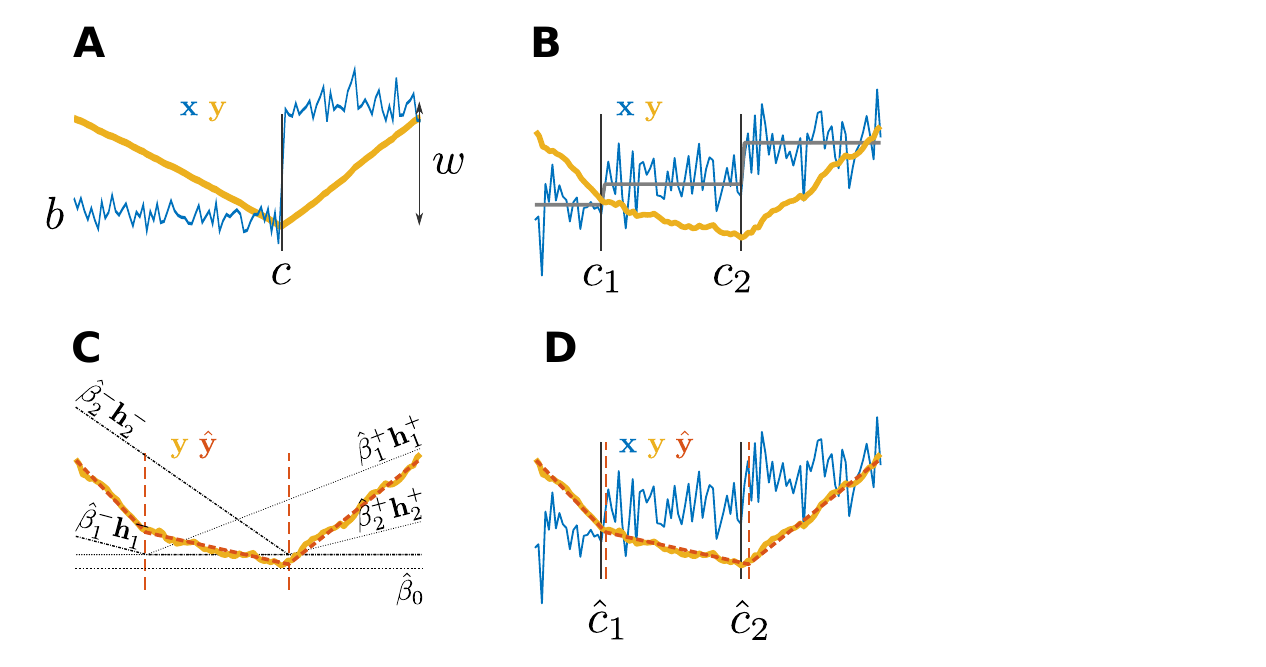}
\caption{Paired Adaptive Regressors for Cumulative Sum; (\textbf{A},\textbf{B}) time series $\mathbf{x}$ with (\textbf{A}) one or (\textbf{B}) two step changes and their corresponding CUSUM transformation $\mathbf{y}$; (\textbf{C}) fitting $\mathbf{y}$ by a piecewise linear model $\mathbf{\hat{y}}$ using two pairs of regressors $\mathbf{h}_1^\pm$ and $\mathbf{h}_2^\pm$; (\textbf{D}) the PARCS model fit $\mathbf{\hat{y}}$ to the CUSUM transformation $\mathbf{y}$ of a time series $\mathbf{x}$, returning estimates of multiple CPs, $\hat{c}_1$ and $\hat{c}_2$.}
\label{fig:01}
\end{figure}

Identifying the presence of a CP requires testing the null hypothesis, $H_0 : w = 0$, against the alternative, $H_1 : w \ne 0$ \citep{Antoch1995, Lombard1994}. This begins by inferring the time of the step according to a CP locator statistic. A typical offline CP locator statistic is the maximum point of the weighted absolute cumulative sum of differences to the mean \citep{Antoch2001, Horvath1997},
\begin{equation} \label{eq:CUSUM}
\hat{c} = \operatorname*{arg\,max}_{0 < t < T}
\biggl( \frac{T}{t(T-t)} \biggr)^\gamma
\Biggl| \sum_{\tau = 1}^t
\bigl( x_\tau - \langle \mathbf{x} \rangle \bigr) \Biggr| ,
\end{equation}
where $\langle \mathbf{x} \rangle$ is the arithmetic mean of the time series (see Figure \ref{fig:01}A). The first term on the right-hand side corrects for bias toward the centre, where more centrally-located points are down-weighed by an amount controlled by parameter $\gamma \in [0,0.5]$. Other CUSUM-based locator statistics exist with different bias-correcting terms and cumulative sum transformations \citep{Antoch1997, Bhattacharya1994, Jirak2012, Kirch2007}. As outlined in the Discussion, PARCS may be modified to include such bias-correcting terms as well. However, as we will demonstrate, PARCS can significantly reduce centre bias even without recourse to such a term. To show this, we will mostly deal with the generic case, $\gamma = 0$, when comparing PARCS to the CUSUM transformation as defined in Eq.~\ref{eq:CUSUM}. This has the added advantage of avoiding having to select an optimal power or an optimal weight factor, a choice that usually depends on prior assumptions on the CP's potential location \citep{Bhattacharya1994}. As such, and unless stated otherwise, the term \emph{CUSUM transformation} will refer, thereof, to the cumulative sum of differences to the mean,
\begin{equation} \label{eq:cusumt}
y_t \triangleq \sum_{\tau = 1}^t
\bigl( x_\tau - \langle \mathbf{x} \rangle \bigr) ,
\end{equation}
where the maximum value,
\begin{equation} \label{eq:Scp}
S = \max_{0 < t < T} \Biggl| \sum_{\tau = 1}^t
\bigl( x_\tau - \langle \mathbf{x} \rangle \bigr) \Biggr| =
\max_{0 < t < T} | y_t | ,
\end{equation}
defines a test statistic by which it is decided whether to reject the null hypothesis.

Given potentially dependent observations, $q > 0$, as defined by the model in Eq.~\ref{eq:model1}, nonparametric bootstrap testing proceeds by \emph{block-permutation} \citep{Davison1997, Huskova2004, Kirch2007}, such that temporal dependence in the data is preserved (see Section \ref{sec:0302}). The candidate CP $\hat{c}$ is identified according to Eq.~\ref{eq:CUSUM} and its associated test statistic $S$ is computed by Eq.~\ref{eq:Scp}. Estimates $\hat{b}$ and $\hat{w}$ are retrieved from the arithmetic means of $\mathbf{x}$ before and after $\hat{c}$ using the model in Eq.~\ref{eq:model1}. By subtracting $\hat{w} \cdot \mathbf{1}_{t-\hat{c}}$ from the time series $\mathbf{x}$ we arrive at a time series $\mathbf{x}_0$ that provides an estimate of the null distribution. The stationary time series $\mathbf{x}_0$ is split into $n$ blocks of size $k$, chosen such that temporal dependencies are mostly preserved in the permuted time series \citep{Davison1997}. One way to do so is to select the block size to be larger than the order of the underlying MA process, $q+1$ \citep[since the autocorrelation function of an MA$(q)$ process cuts off at order $q$;][]{Davison1997}. This requires identifying the order $q$ which can be determined from the $H_0$-conform time series $\mathbf{x}_0$ by inspecting its autocorrelation function \citep{Fan2003} for different time lags $\tau$. The autocorrelation function's asymptotic distribution \citep{Kendall1983},
\begin{equation} \label{eq:acorr}
\mathrm{acorr} (\mathbf{x}_0 ; \tau) \sim \mathcal{N} \bigl( -1/(T-\tau) , 
+1/(T-\tau) \bigr) ,
\end{equation}
provides a test statistic for deciding the largest time lag $q$ at which to reject the null hypothesis $H_0:\mathrm{acorr} (\mathbf{x}_0 ; q) = 0$, given some preset significance level $\alpha \in [0 , 1]$. The resulting blocks are randomly permuted and each permutation is CUSUM-transformed according to Eq.~\ref{eq:cusumt} to compute an $H_0$-conform sample $S_0$ of the test statistic $S$ in Eq.~\ref{eq:Scp} (note that we do not know the true step parameter $w$ or the true CP $c$, of course, such that this procedure will only yield an estimate of the $H_0$ distribution). A sufficiently large number $B$ of permutations results in samples $S_i$ of an $H_0$-conform \emph{empirical distribution function} (EDF) $\mathrm{F}(S_0) \triangleq \sum_{i=1}^B \mathbf{1}_{S_0 - S_i} / B$ that weighs every sample $S_i$ equally. The candidate CP $\hat{c}$ is detected when the test statistic $S$ as computed from the original time series $\mathbf{x}$ satisfies $S \ge \mathrm{F}^{-1} (1 - \alpha)$, where $\alpha$ is a preset significance level and $\mathrm{F}^{-1}(1-\alpha)$ the inverse of the EDF, defined as the $(1-\alpha) B^\mathrm{th}$ largest value out of $B$ permutations \citep{Davison1997, Durstewitz2017}.

\subsection{PARCS: Paired Adaptive Regressors for Cumulative Sum} \label{sec:0202}

The PARCS method for estimating multiple CPs rests on the fact that the integral of a piecewise constant function is piecewise linear. The AMOC model as defined in Eq.~\ref{eq:model1} assumes a piecewise stationary MA process, consisting of two segments with constant mean. A process consisting of $M+1$ segments generalises Eq.~\ref{eq:model1} to data containing at-most-$M$-change,
\begin{equation} \label{eq:xM}
x_t = b + \sum_{m=1}^M w_m \cdot \mathbf{1}_{t-c_m} +
\sum_{\tau \ge 0} \kappa_{\tau} \epsilon_{t-\tau}
\quad ; \quad \kappa_0 = 1 , \;
\epsilon_t \sim \mathcal{N} (0,\sigma^2) .
\end{equation}

The CUSUM transformation $\mathbf{y} = \{y_t\}_{1:T}$ of this process as given by Eq.~\ref{eq:cusumt} corresponds to the numerical integration of a piecewise stationary process $\mathbf{x} - \langle \mathbf{x} \rangle$. That is, $\mathbf{y}$ is approximately (due to the noise) piecewise linear (exactly piecewise linear \emph{in the mean}; see Figure \ref{fig:01}B). If points $\{c_m\}_{1:M}$ at which $\mathbf{y}$ bends were known, the latter can be fitted by a weighted sum of local piecewise linear basis functions or \emph{splines}, centred at the \emph{knots} $\{c_m\}_{1:M}$,
\begin{equation*} 
h_{t,c_m}^+ = 
\begin{cases}
t - c_m \quad & \mathrm{if} \ t > c_m \\
0 \quad & \mathrm{otherwise}
\end{cases} \quad \mathrm{and} \quad
h_{t,c_m}^- = 
\begin{cases}
c_m - t \quad & \mathrm{if} \ t < c_m \\
0 \quad & \mathrm{otherwise}
\end{cases} \quad .
\end{equation*}

This fit corresponds to modelling the expected value of $\mathbf{y}$, conditioned on spline pair set $\mathcal{H} = \{ \mathbf{h}_{c_m}^\pm \}_{1:M}$, resulting in model inference,
\begin{equation*}
\mathrm{E} \bigl[ y_t \bigl| \mathcal{H} \bigr] \approx
\hat{y}_{t,M} = \hat{\beta}_0 + \sum_{m = 1}^M \hat{\beta}_m^+ h_{t,c_m}^+ +
\sum_{m = 1}^M \hat{\beta}_m^- h_{t,c_m}^- ,
\end{equation*}
which is a simple regression problem that can be solved by estimating the intercept $\hat{\beta}_0$ and coefficients $\hat{\beta}_m^{\pm}$ that minimise the mean-square-error,
\begin{equation} \label{eq:mse}
\mathrm{mse}_M (\mathbf{y}, \hat{\mathbf{y}}_M) =
\frac{1}{T} \sum_{t=1}^T (y_t - \hat{y}_{t,M})^2 .
\end{equation}

However, in the multiple CP detection setting (assuming $M$ is known), optimal knot placement is not known \emph{a priori}, but can be inferred by \emph{adaptively} adjusting knot locations \citep{Friedman1991, Friedman1989, Stone1997} to maximally satisfy the goodness-of-fit criterion in Eq.~\ref{eq:mse}. In other words, and as shown in Figures \ref{fig:01}C,D, the problem of identifying multiple CPs is replaced by the equivalent problem of inferring the order-$M$ PARCS model (or PARCS$_M$ model),
\begin{equation} \label{eq:parcs}
\hat{\mathbf{y}}_M = \hat{\beta}_0 +
\sum_{m=1}^M \hat{\beta}_m^+ \mathbf{h}_{\hat{c}_m}^+ +
\sum_{m=1}^M \hat{\beta}_m^- \mathbf{h}_{\hat{c}_m}^- ,
\end{equation}
with associated $M$-tuple $\hat{\mathbf{c}} \triangleq (\hat{c}_m)_{1:M}$ that best fits the CUSUM transformation of the time series. Regression coefficients in model \ref{eq:parcs} are real numbers, while knots in the present time series context are positive integers, excluding the first and last time steps, $\hat{c}_m \in \{2,3,\dots,T-1\}$.

Fitting the PARCS$_M$ model is based on a forward/backward spline selection strategy \citep{Smith1982} with added CP ranking stage and proceeds as outlined in Algorithm \ref{alg:01}. Starting with an empty PARCS$_0$ model, containing only the intercept $\hat{\beta}_0$, a \emph{forward} sweep increases model complexity to a forward upper bound order $L > M$ by adding at each iteration the spline pair $\mathbf{h}_c^{\pm}$, not yet contained in the model, that decreases residual mean-square-error the most. A reasonable heuristic for setting $L$ is 2 to 3 times $M$ (assuming $M$ is known or given some liberal guess). This is followed by a backward \emph{pruning} iteration, in which the spline pair whose removal increases residual mean-square-error the least is dropped from the model. Pruning removes those knots that were added at the beginning of the forward phase which became redundant as the model was refined by later additions \citep{Friedman1989}. This stage continues until the number of knots reaches the preset final upper bound of model complexity $M$, i.e., $L-M$ knots are pruned. Knots are then sorted in descending order according to the amount of explained variance. The \emph{ranking} iteration returns a nested model by pruning the PARCS$_M$ model further, down to the PARCS$_0$ model. The first knot to be pruned, reducing the number of knots to $M-1$, explains the least variance and is placed last as $\hat{c}_M$ in the $M$-tuple $\hat{\mathbf{c}}$. The last knot to be pruned explains the most variance and is placed first as $\hat{c}_1$. Note that regression coefficients are re-estimated every time a knot is added to or removed from the PARCS model.

\begin{algorithm}[!t]
\DontPrintSemicolon
\SetAlgoNoEnd
\SetAlgoNoLine
\KwIn{$L, M \ \mathrm{and} \ \mathbf{y}$}
\KwOut{$\hat{\mathbf{c}} \triangleq (\hat{c}_m)_{1:M} \ 
\mathrm{and} \ \hat{\mathbf{y}}_M$}
$\hat{\mathbf{c}} , \mathcal{H} \leftarrow \varnothing$ \; 
\For(\tcp*[f]{forward stage}){$m \leftarrow 1$ \KwTo $L$}
{$\hat{c}_m \leftarrow \operatorname*{arg\,min}_{1 < c < T} \bigl\{ 
\mathrm{mse}_m(\mathbf{y}, \hat{\mathbf{y}}_{\mathcal{H} \cup 
\mathbf{h}_c^{\pm}}) \bigl| 
\mathbf{h}_c^{\pm} \notin \mathcal{H}\bigr\}$ \;
$\mathcal{H} \leftarrow \mathcal{H} \cup \mathbf{h}_{\hat{c}_m}^{\pm}$ \;
}
\For(\tcp*[f]{pruning stage}){$m \leftarrow L$ \KwTo $M+1$}
{$\hat{c} \leftarrow \operatorname*{arg\,min}_c \bigl\{ 
\mathrm{mse}_{m-1}(\mathbf{y}, \hat{\mathbf{y}}_{\mathcal{H} \backslash 
\mathbf{h}_c^{\pm}}) \bigr\}$ \;
$\hat{\mathbf{c}} \leftarrow \hat{\mathbf{c}} \backslash \hat{c}
\quad \mathrm{and} \quad \mathcal{H} \leftarrow
\mathcal{H} \backslash \mathbf{h}_{\hat{c}}^\pm$ \;
}
$\hat{\mathbf{y}}_M \leftarrow \hat{\beta}_0 +
\sum_{m=1}^M \hat{\beta}_m^+ \mathbf{h}_{\hat{c}_m}^+ +
\sum_{m=1}^M \hat{\beta}_m^- \mathbf{h}_{\hat{c}_m}^-$ \;
\For(\tcp*[f]{ranking stage}){$m \leftarrow M$ \KwTo $1$}
{$\hat{c}_m \leftarrow \operatorname*{arg\,min}_c \bigl\{ 
\mathrm{mse}_{m-1}(\mathbf{y}, \hat{\mathbf{y}}_{\mathcal{H} \backslash 
\mathbf{h}_c^{\pm}}) \bigr\}$ \;
$\mathcal{H} \leftarrow \mathcal{H} \backslash \mathbf{h}_{\hat{c}_m}^{\pm}$ \;
}
\caption{Procedure for inferring the PARCS$_M$ model with forward/backward spline selection (first/second loop) and CP ranking (third loop). Regression coefficients are computed by least squares estimation, conditioned on the set of knot locations of predefined size $M$ that minimises mean-square-error. Final knot locations are specified by eliminating spurious knots through block-permutation bootstrapping as described in Section \ref{sec:0203}.} \label{alg:01}
\end{algorithm}

The model can be effortlessly extended to the multiple response setting in the case of spatially independent time series \citep[extension to a nondiagonal MA covariance matrix,][will be considered elsewhere]{Stone1997}. Given $N$ independent, piecewise stationary MA processes with common CPs $\{c_m\}_{1:M}$, 
\begin{equation} \label{eq:xMN}
x_{t,n} = b_n + \sum_{m=1}^M w_{mn} \cdot \mathbf{1}_{t-c_m} +
\sum_{\tau \ge 0} \kappa_\tau \epsilon_{t-\tau,n}
\quad ; \quad \kappa_0 = 1 , \; \epsilon_{t,n} \sim \mathcal{N} (0,\sigma^2) ,
\end{equation}
where $n = 1, \dots , N$, the corresponding multivariate CUSUM transformation $\mathbf{y}_t = \{y_{t,n}\}_{1:N}$ is fitted by the multiple response, PARCS$_M$ model, conditioned on common spline pairs,
\begin{equation*} \label{eq:parcsN}
\mathrm{E} \bigl[ y_{t,n} \bigl| \mathcal{H} \bigr] \approx
\hat{y}_{t,M,n} = \hat{\beta}_{0n} +
\sum_{m = 1}^M \hat{\beta}_{mn}^+ h_{t,\hat{c}_m}^+ +
\sum_{m = 1}^M \hat{\beta}_{mn}^- h_{t,\hat{c}_m}^- ,
\end{equation*}
using Algorithm \ref{alg:01}. Returning CPs that are common to all variables $\mathbf{x}_n$ is done by using the goodness-of-fit criterion in Eq.~\ref{eq:mse}, averaged over all responses $\mathbf{y}_n$.

\subsection{PARCS Model Selection by Block-Permutation Bootstrap} \label{sec:0203}

The piecewise linear PARCS formulation, Eq.~\ref{eq:parcs}, of the CUSUM transformation in Eq.~\ref{eq:cusumt} bends at the CPs. Due to the presence of noise in the original time series $\mathbf{x}$, some noise realisations may appear as slight bends in the CUSUM-transformed time series, leading PARCS to return false CPs. As such, the amount of bending at knot $\hat{c}_m$ can be used as a test statistic for bootstrap significance testing that can refine the PARCS model further. No bending indicates either a constant fit, $\hat{\beta}_m^+ = \hat{\beta}_m^- = 0$, or a smooth linear fit, $\hat{\beta}_m^+ = - \hat{\beta}_m^-$ (see also Figure \ref{fig:01}C). Thus, a suitable test statistic that quantifies the amount of bending at $\hat{c}_m$ is given by,
\begin{equation} \label{eq:Sparcs}
S = \bigl| \hat{\beta}_m^+ + \hat{\beta}_m^- \bigr| ,
\end{equation}
where for multivariate time series, the test statistic is the average over all time series.

Before describing the block-permutation bootstrap method for PARCS, we outline a procedure for identifying the order $q$ of the MA noise process, provided as pseudocode in Algorithm \ref{alg:02}. First, an $H_0$-conform time series $\mathbf{x}_0 = \{x_{t,0}\}_{1:T}$ is computed by regressing out the PARCS model $\hat{\mathbf{y}}_M$ of Eq.~\ref{eq:parcs} from the CUSUM-transformed time series $\mathbf{y}$ and then inverting the CUSUM transformation. This is followed by inspecting the autocorrelation function of $\mathbf{x}_0$ for different time lags $\tau$. The largest time lag at which the null hypothesis $H_0 : \mathrm{acorr} (\mathbf{x}_0 ; q) = 0$ is rejected, given some preset significance level $\alpha \in [0 , 1]$ is then returned as the order $q$, given some predefined upper bound of MA order, $Q$.

\begin{algorithm}[!t]
\DontPrintSemicolon
\SetAlgoNoEnd
\SetAlgoNoLine
\KwIn{$\mathbf{y}, \hat{\mathbf{c}}, Q \ \mathrm{and} \ \alpha$}
\KwOut{$q \le Q $}
$q \leftarrow Q$\;
$\mathbf{y}_0 \leftarrow \mathbf{y} - \Bigl( \hat{\beta}_0 +
\sum_{m=1}^M \hat{\beta}_m^+ \mathbf{h}_{\hat{c}_m}^+ +
\sum_{m=1}^M \hat{\beta}_m^- \mathbf{h}_{\hat{c}_m}^- \Bigr)$ \;
$x_{t,0} \leftarrow y_{t,0} - y_{t-1,0} + \langle \mathbf{x} \rangle \quad 
\mathrm{for} \quad t = 1,\dots,T \quad \mathrm{where} \quad y_{0,0} = 0$ \;

\For{$\tau \leftarrow 1$ \KwTo $Q$}
{$\mathrm{F}_\tau \leftarrow \mathrm{CDF} \Bigl( \mathcal{N} \bigl( -1/(T-\tau) , 
+1/(T-\tau) \bigr) \Bigr)$ \;

\If{$\mathrm{acorr} (\mathbf{x}_0 ; \tau) \in \bigl[\mathrm{F}_\tau^{-1} (\alpha/2) , \mathrm{F}_\tau^{-1} (1 - \alpha/2) \bigr]$}
{$q \leftarrow \tau - 1$ \;
\textbf{break}\;}
}
\caption{Identifying the order $q$ of the MA process, given some upper bound $Q$. The $H_0$-conform time series $\mathbf{x}_0$ is estimated before entering the loop. The loop increases the autocorrelation time lag and exits when the autocorrelation of $\mathbf{x}_0$ is not significantly different from 0 anymore.} \label{alg:02}
\end{algorithm}

Given the $M$-tuple CP set $\hat{\mathbf{c}}$ returned by Algorithm \ref{alg:01} and an estimate of the dependent normal noise order $q$ by Algorithm \ref{alg:02}, a block-permutation bootstrap test returns the subset $\hat{\bm{\varsigma}}$ of significant CPs, as outlined in Algorithm \ref{alg:03}. First, an $H_0$-conform time series $\mathbf{x}_0$ is computed. For each CP $\hat{c}_m \in \hat{\mathbf{c}}$, starting with the one ranked highest, all CP-splines already deemed significant by the bootstrap test are regressed out of $\mathbf{y}$. A PARCS model with the remaining knots, including $\hat{c}_m$, is estimated and the test statistic $S$, evaluated at $\hat{c}_m$ according to Eq.~\ref{eq:Sparcs}, is computed. Knot $\hat{c}_m$ is tested for significance against an $H_0$-conform EDF, estimated through block-permutation bootstrapping: A total of $B$ bootstrap samples is generated from the $H_0$-conform series $\mathbf{x}_0$ by randomly permuting blocks of size $k=q+1$. For each of these $i = 1 \dots B$ bootstrap samples test statistic $S_i$ is evaluated at knot location $\hat{c}_m$, yielding an EDF $\mathrm{F}(S_0)$ which assigns equal probability $1/B$ to each bootstrapped $S_i$. A significant $\hat{c}_m$ is then added to $\hat{\bm{\varsigma}}$, or rejected as false discovery otherwise. The procedure repeats for the CP next in the rank order. Similar to Algorithm \ref{alg:01}, regression coefficients are re-estimated every time a knot is added to or removed from the PARCS model.

\begin{algorithm}[!ht]
\DontPrintSemicolon
\SetAlgoNoEnd
\SetAlgoNoLine
\KwIn{$\mathbf{y}, \hat{\mathbf{c}}, k, B \ \mathrm{and} \ \alpha$}
\KwOut{$\hat{\bm{\varsigma}} \subseteq \hat{\mathbf{c}}$}
$\hat{\bm{\varsigma}} \leftarrow \varnothing$ \;
$\mathbf{y}_0 \leftarrow \mathbf{y} - \Bigl( \hat{\beta}_0 +
\sum_{m=1}^M \hat{\beta}_m^+ \mathbf{h}_{\hat{c}_m}^+ +
\sum_{m=1}^M \hat{\beta}_m^- \mathbf{h}_{\hat{c}_m}^- \Bigr)$ \;
$x_{t,0} \leftarrow y_{t,0} - y_{t-1,0} + \langle \mathbf{x} \rangle \quad 
\mathrm{for} \quad t = 1,\dots,T \quad \mathrm{where} \quad y_{0,0} = 0$ \;

\For{$m \leftarrow 1$ \KwTo $M$}
{$\mathbf{y}_{\hat{\mathbf{c}} \backslash \hat{\bm{\varsigma}}}
\leftarrow \mathbf{y} - \Bigl( \hat{\beta}_0 +
\sum_{\mu = 1}^{|\hat{\bm{\varsigma}}|} \hat{\beta}_\mu^+
\mathbf{h}_{\hat{\varsigma}_\mu}^+ +
\sum_{\mu = 1}^{|\hat{\bm{\varsigma}}|} \hat{\beta}_\mu^-
\mathbf{h}_{\hat{\varsigma}_\mu}^- \Bigr)$ \;
$\hat{\mathbf{y}}_{\hat{\mathbf{c}} \backslash \hat{\bm{\varsigma}}}
\leftarrow \hat{\beta}_0 +
\sum_{\mu = m}^M \hat{\beta}_\mu^+ \mathbf{h}_{\hat{c}_\mu}^+ +
\sum_{\mu = m}^M \hat{\beta}_\mu^- \mathbf{h}_{\hat{c}_\mu}^-$ \;
$S \leftarrow \bigl| \hat{\beta}_m^+ + \hat{\beta}_m^- \bigr|$ \;
$\mathrm{F}(S_0) \leftarrow \mathrm{BlockPermutationBootstrap} (\mathbf{x}_0,\hat{c}_m,k,B)$ \;
\If{$S \ge \mathrm{F}^{-1} (1 - \alpha)$}
{$\hat{\bm{\varsigma}} \leftarrow \hat{\bm{\varsigma}} \cup \hat{c}_m$}
}
\caption{Block-permutation bootstrap procedure for PARCS, given block size $k$. The $H_0$-conform time series $\mathbf{x}_0$ is estimated before entering the loop. The loop iterates over the rank-ordered CPs to test for each CP's significance.} \label{alg:03}
\end{algorithm}

\section{Results} \label{sec:03}

We first evaluate the PARCS method on synthetic data in single and multiple CP detection settings, followed by a real data example on detecting behavioural and neural change points during rule learning.

\subsection{Alleviating CUSUM Bias in AMOC Detection} \label{sec:0301}

We first compare the CUSUM method for detecting a single CP to the PARCS approach in order to evaluate the effect of each method on the centre bias in CP detection. Both white and MA Gaussian noise are considered. We also compare PARCS to the CUSUM locator statistic of Eq.~\ref{eq:CUSUM} with $\gamma = 0.5$ (the maximum likelihood estimator of CP location under the assumption of i.i.d.\ Gaussian noise) and identify conditions under which one method is preferable over the other.

Univariate time series of length $T = 100$ are simulated according to the step model in Eq.~\ref{eq:model1} with different levels of white Gaussian noise, $\sigma \in \{ 0.4 , 0.5 , \dots , 1.0 \}$, and different ground truth CP locations, $c \in \{ 20 , 30 , \dots , 80 \}$. Baseline is set to $b = 0$ and step parameter to $w = 1$. Note that in the step model with white Gaussian noise, increasing $\sigma$ is equivalent to reducing $w$. A single CP was identified by using the CUSUM method and estimating the PARCS$_1$ model, both followed by bootstrap significance testing with $B = 10000$ permutations, nominal significance level $\alpha = 0.05$, and blocks of size $k = 1$ (since noise is independent in this example). Each parameter configuration was repeated for 1000 noise realisations.

\begin{figure}[p]
\centering
\includegraphics[width=\linewidth]{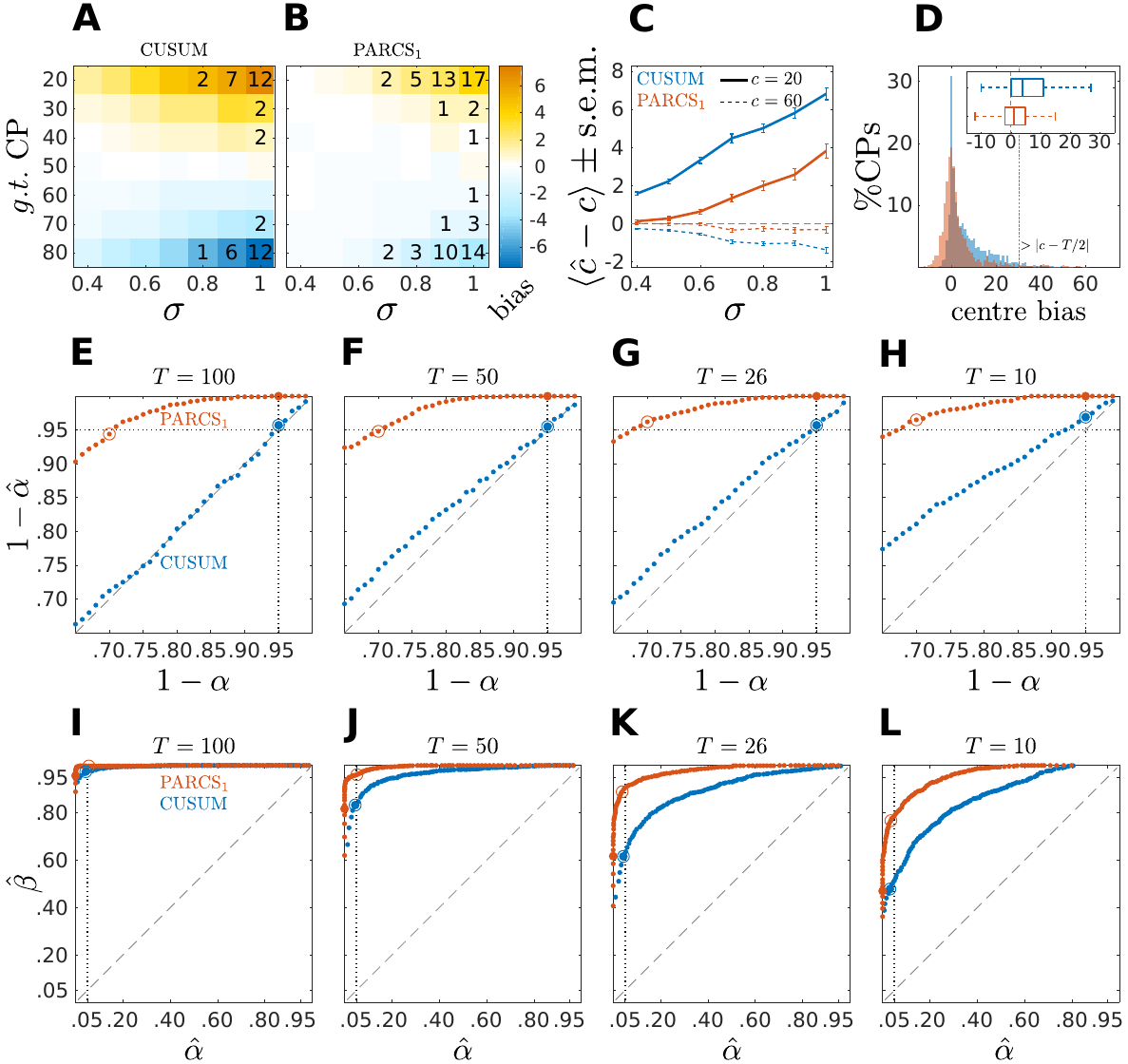}
\caption{Centre bias in PARCS compared to CUSUM for temporally independent noise; (\textbf{A},\textbf{B}) bias, $\langle \hat{c} - c \rangle$, colour-coded as indicated by the colour bar; numbers indicate rounded type II error rates; (\textbf{C}) bias $\pm$ s.e.m.\ for $c = 20$ (solid) and $c = 60$ (dashed); (\textbf{D}) centre bias distributions for $c \in \{20,80\}$ and $\sigma = 1.0$; inset shows centre bias distributions as boxplots that mark the median and first and third quartiles; whiskers include points within 1.5 times the interquartile range; outliers are excluded; (\textbf{E}-\textbf{H}) P-P plots comparing nominal ($x$-axis) versus factual ($y$-axis) true $H_0$ rejection rates in time series of length (E) $T = 100$, (F) $T = 50$, (G) $T = 26$, and (H) $T = 10$; dotted vertical line, nominal $\alpha = 0.05$; dotted horizontal line, factual $\hat{\alpha} = 0.05$; (\textbf{I}-\textbf{L}) ROC curves depicting false discovery rate (type I error rate; $x$-axis) versus power ($y$-axis) for different series lengths as in E-H; dotted vertical line, nominal $\alpha = 0.05$; In E-L, larger filled circles indicate the empirical $H_0$ rejection rates at a nominal $\alpha = 0.05$, and empty circles indicate where the factual $\hat{\alpha} \approx 0.05$.}
\label{fig:02}
\end{figure}

We compare bias in CP detection toward the centre of the time series in both the CUSUM and PARCS methods. We measure this \emph{centre bias} by $\mathrm{cb} = (2 \cdot \mathbf{1}_{c - T/2} -1) \cdot (c - \hat{c})$, which is positive when estimate $\hat{c}$ falls onto the side located toward the centre from $c$, and is negative otherwise. As expected given the choice $\gamma = 0$ in Eq.~\ref{eq:CUSUM}, the CUSUM method shows a strong centre bias which increases for lower signal-to-noise ratio and more peripheral CPs (see Figure \ref{fig:02}A). The CUSUM method's power decreases for harder parameter settings (higher $\sigma$ and more peripheral $c$) in that true CPs are missed in more of the realisations. The PARCS method results in a significant reduction in centre bias but does not eliminate it completely, and yields more misses relative to CUSUM if both are run at the same nominal $\alpha$ level (see Figure \ref{fig:02}B). Summary comparison between the two methods is shown in Figure \ref{fig:02}C for two exemplary CP locations, $c \in \{ 20 , 60 \}$.

To fully appreciate the source of CUSUM centre bias and its reduction by PARCS, time series realisations with the two hardest parameter settings ($\sigma = 1.0$ and $c \in \{ 20 , 80 \}$) are considered in Figure \ref{fig:02}D, which compares the distribution of $\mathrm{cb}$ in the $81\%$ of realisations in which both CUSUM and PARCS returned a CP. The histograms show a strongly skewed, heavy-tailed distribution for CUSUM, compared to a more symmetric distribution around 0 for PARCS, indicating only little bias. Most of the centre bias in PARCS is accounted for by outliers. This is illustrated by excluding outliers in the boxplots, which show a median of 1 time step centre bias in PARCS against median centre bias of 4 time steps in the case of CUSUM. Note that measuring centre bias as defined above does not differentiate between biased detections and false discoveries where, in extreme cases, a CP may be detected beyond the middle point $T/2$ of the time series, corresponding to centre bias greater than $|c - T/2|$. However, this scenario rarely occurred in the simulation results reported here.

While PARCS reduces centre bias, the simulation results above indicate that it behaves more conservatively than CUSUM at the same nominal $\alpha$ level. In principle, false $H_1$ rejection rates (type II errors) may be reduced by adjusting the $\alpha$ level, at the same time producing more false discoveries. In order to assess how well the nominal significance level $\alpha$ agrees with the empirical type I error rate (false discoveries), 1000 white Gaussian noise realisations of length $T = 100$ are simulated with $\sigma = 1.0$. Conclusions drawn from this analysis are largely the same for larger signal-to-noise ratio (results not shown). A single CP was extracted using the CUSUM method and estimating the PARCS$_1$ model on these time series conforming to the null hypothesis $H_0 : w = 0$. Type I error rates at different nominal $\alpha$ levels are shown in Figure \ref{fig:02}E as \emph{probability-probability} (P-P) plots, depicting the nominal, $1-\alpha$, against the empirical, $1-\hat{\alpha}$, probabilities of accepting the null hypothesis when the null hypothesis is true. While the empirical type I error rate of CUSUM perfectly agrees with the nominal significance level, for PARCS, in contrast, the empirical rate of false discoveries tends toward 0\% for $\alpha = 0.05$ and remains smaller than 1\% for $\alpha$ as large as $0.18$. This entails that PARCS behaves highly conservatively, and that the nominal $\alpha$ level may be adjusted considerably upward without strongly influencing the false discovery rate. On the other hand, despite being more conservative, Figure \ref{fig:02}I shows that the \emph{receiver operating characteristic} (ROC) curve for PARCS, depicting the method's false discovery rate against its power for different nominal $\alpha$ levels, consistently lies above that of CUSUM. For estimating the statistical power of each method, 1000 white Gaussian noise realisations with one CP at a random location in the range $[20,80]\%T$ (and $w = \sigma = 1$) are simulated and type II error rates at different nominal $\alpha$ levels are computed. This ROC analysis indicates that for every nominal $\alpha$ level for CUSUM, there exists at least one nominal $\alpha$ for PARCS such that PARCS has both higher power (fewer type II errors) and lower false detection rate (fewer type I errors), making it the preferable method. This point is explored in more detail later in the context of multiple CP detection in Section \ref{sec:0302}. For shorter time series, PARCS behaves similarly as for longer time series in our simulations, while for CUSUM type I error rates now start to fall below the nominal $\alpha$ level as well (Figures \ref{fig:02}F-H). The area under the ROC curves become smaller for shorter time series for both methods, but the ROC curve of PARCS consistently lies above that of CUSUM in those cases as well (Figures \ref{fig:02}J-L).

Next, we examine the behaviour of the tests with dependent noise. In the case of temporally dependent noise, an appropriate block size for the bootstrap procedure could be determined by inspecting the autocorrelation function of $\mathbf{x}_0$ (see Eq.~\ref{eq:acorr} and Algorithm \ref{alg:02}). 1000 noise realisations of length $T=100$ are drawn from an order-2 MA process with coefficients $\kappa_1 = -0.5 / \sigma$ and $\kappa_2 = 0.4 / \sigma$. Since increasing noise variance in the temporally dependent case is not equivalent to decreasing the step parameter, we repeat the analysis with the same parameters from the temporally independent case above but varying the step parameter, $w \in \{ 0.7 , 0.8 , \dots , 1.3 \}$ and considering two levels of Gaussian noise, $\sigma \in \{0.7 , 1.0\}$.

\begin{figure}[!t]
\centering
\includegraphics[width=\linewidth]{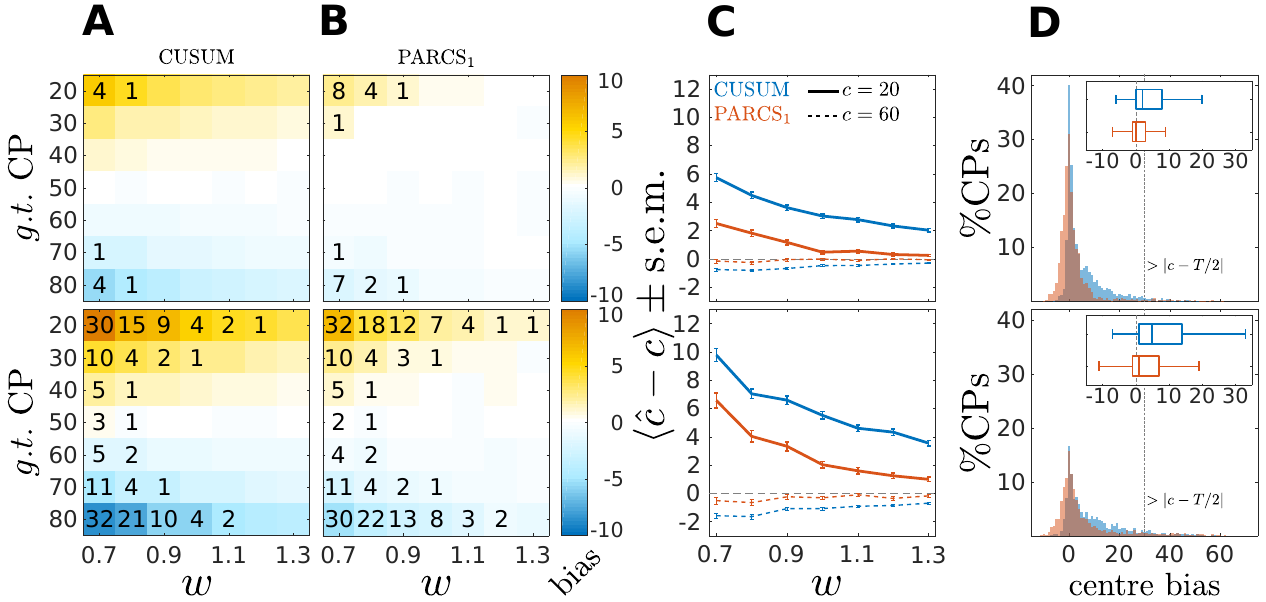}
\caption{Centre bias in PARCS compared to CUSUM for temporally dependent noise with $\sigma = 0.7$ (top) and $\sigma = 1.0$ (bottom); (\textbf{A},\textbf{B}) bias, $\langle \hat{c} - c \rangle$, colour-coded as indicated by the colour bar; numbers indicate rounded rounded type II error rates; (\textbf{C}) bias $\pm$ s.e.m.\ for $c = 20$ (solid) and $c = 60$ (dashed); (\textbf{D}) centre bias distributions for $c \in \{20,80\}$ and $\sigma = 1.0$; inset shows centre bias distributions as boxplots that mark the median and first and third quartiles; whiskers include points within 1.5 times the interquartile range; outliers are excluded.}
\label{fig:03}
\end{figure}

Figure \ref{fig:03} shows results of the comparison for $\sigma = 0.7$ (top row) and $\sigma = 1.0$ (bottom row). Similar to the white noise case, the CUSUM method's centre bias increases for smaller signal-to-noise ratio (smaller $w$), and PARCS, in comparison, consistently reduces centre bias. For the same nominal $\alpha$ level, PARCS misses more of the true CPs than CUSUM for peripheral CPs when $\sigma = 0.7$ (top panels of Figures \ref{fig:03}A,B), but the type II error rate￹ of the two methods is more comparable in the high noise case, $\sigma = 1.0$ (bottom panels of Figures \ref{fig:03}A,B), despite the PARCS method's more conservative behaviour (far lower type I error rates) in this setting as well￹. A summary comparison between the two methods is shown in Figure \ref{fig:03}C for two exemplary CP locations, $c \in \{ 20 , 60 \}$. Despite a significant reduction when using PARCS, both centre bias distributions in the 62\% of realisations with a CP identified by the two methods, and with the two hardest parameter settings ($c \in \{ 20 , 80 \}$, $\sigma = 1.0$ and $w = 0.7$) remain strongly skewed (bottom panel of Figure \ref{fig:03}D).

So far, we compared PARCS to the CUSUM statistic with $\gamma = 0$. It is intuitive when developing PARCS to choose $\gamma = 0$ for the CUSUM transformation in Eq.~\ref{eq:CUSUM}, since this directly corresponds to the numerical integral of the time series upon which the PARCS approach is based (but see Section \ref{sec:04}). Besides, under certain conditions, the CUSUM method using the test statistic with $\gamma < 0.5$ is more sensitive than that with $\gamma = 0.5$ \citep{Antoch1995}. However, the CUSUM statistic with $\gamma = 0.5$ returns the maximum likelihood estimator of CP location in an AMOC scenario when noise in the step model of Eq.~\ref{eq:model1} is i.i.d.\ and normally distributed, leading theoretically to the strongest centre bias reduction under those conditions \citep{Antoch2001}. We therefore also compare PARCS to this maximum likelihood CUSUM estimator here, henceforth referred to as CUSUM$_\text{ML}$.

\begin{figure}[!t]
\centering
\includegraphics[width=\linewidth]{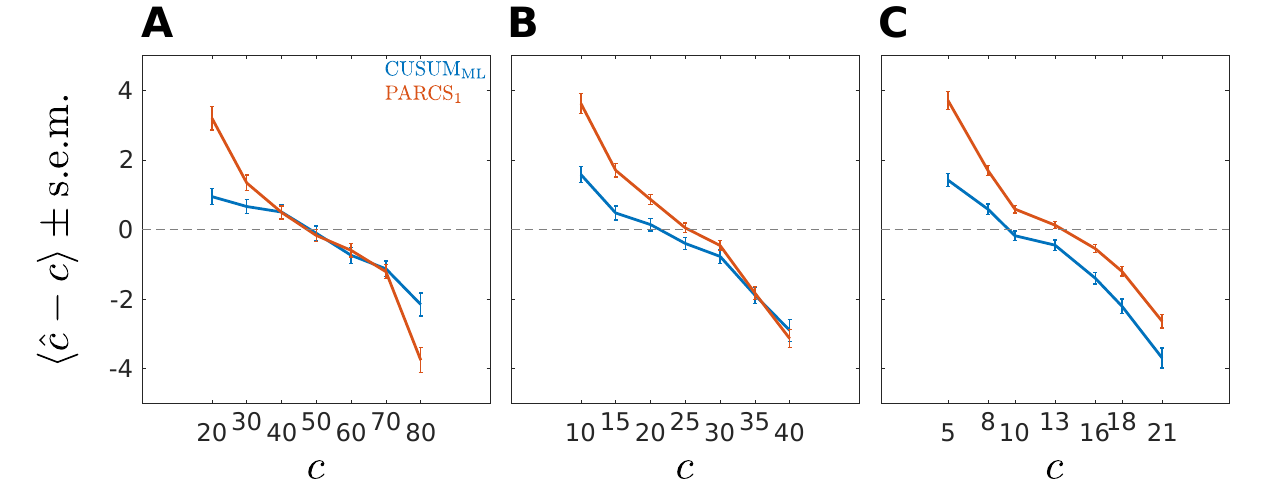}
\caption{Bias $\pm$ s.e.m.\ in PARCS compared to CUSUM$_\text{ML}$ with time series of length (\textbf{A}) $T = 100$, (\textbf{B}) $T = 50$, and (\textbf{C}) $T = 26$; noise is temporally independent with $\sigma = 1.0$.}
\label{fig:04}
\end{figure}

Univerariate time series of length $T = 100$ are simulated according to the step model Eq.~\ref{eq:model1} with different ground truth CP locations, $c \in \{ 20 , 30 , \dots , 80 \}$. We consider only the scenario with largest white Gaussian noise variance, $\sigma = 1.0$, in this analysis, for which PARCS showed the largest centre bias. A single CP was identified by using the CUSUM$_\text{ML}$ method and estimating the PARCS$_1$ model, both followed by bootstrap significance testing. Other parameters are as in the previous analyses above. CUSUM$_\text{ML}$ results in a significant reduction in centre bias compared to PARCS in three of the most peripheral ground truth CPs, $c \in \{20,30,80\}$, but does not eliminate it completely (see Figure \ref{fig:04}A). For the same nominal $\alpha$ level, CUSUM$_\text{ML}$ also shows lower type II error rates compared to PARCS for these CPs as reported in Table \ref{tab:01}, but recall that PARCS has a far lower type I error rate than CUSUM for the same choice of nominal $\alpha$ \citep[cf.\ Figures \ref{fig:02}E-H and][]{Kirch2007}. The two methods are comparable in the quality of their detections for all other ground truth CP locations, $c \in \{40,\dots,70\}$, with PARCS having a slight advantage.

In order to assess how well the two methods fare in the small sample size limit, and to characterise the convergence behaviour of the bootstrap procedure in each method, we repeat the same analysis for shorter series lengths, $T \in \{ 50, 26\}$. Ground truth CPs are set to the same relative location within the time series as in the $T = 100$ simulations. As summarised in Table \ref{tab:01}, detection rates deteriorate as series length decreases, as does the bias relative to series length (where the relative location within the series with respect to the periphery is more relevant than the absolute CP location; see Figures \ref{fig:04}B,C). Especially for $T = 26$, PARCS performs mostly better than CUSUM$_\text{ML}$, giving higher detection rates (see Table \ref{tab:01}) and smaller centre bias (Figure \ref{fig:04}C) in the majority of ground truth CPs, although it is still more conservative with near 0\% type I error rate (given the bootstrap resolution; see Figure \ref{fig:02}G). As we show next, this is a particularly important advantage of PARCS over the CUSUM-based methods when detecting multiple CPs, since CUSUM-based techniques rely on dissecting the time series into smaller segments in this case, reducing sample size at each iteration.

\begin{table}[!t]
\centering
\begin{tabular}{c | c | c c c c c c c c c}
$T$ & \textbf{method} & & \multicolumn{7}{c}{\textbf{type II error rate}} \\
& & $c = \mathrm{round}($ & $20$ & $30$ & $40$ & $50$ & $60$ & $70$ & $80$ & $\%T)$ \\
\hline \hline
& CUSUM$_\text{ML}$ & & \underline{07} & \underline{03} & 02 & 01 & 01 & \underline{03} & \underline{12} & \\
100 & & & & & & & & & & \\
& PARCS & & 17 & 03 & \underline{01} & \underline{01} & \underline{01} & 03 & 16 & \\
\hline
& CUSUM$_\text{ML}$ & & \underline{30} & \underline{20} & 16 & 13 & 16 & 26 & \underline{37} & \\
50 & & & & & & & & & & \\
& PARCS & & 44 & 22 & \underline{12} & \underline{08} & \underline{10} & \underline{19} & 41 & \\
\hline
& CUSUM$_\text{ML}$ & & \underline{53} & \underline{38} & 35 & 32 & 38 & 44 & 59 & \\
26 & & & & & & & & & & \\
& PARCS & & 68 & 41 & \underline{32} & \underline{24} & \underline{29} & \underline{37} & \underline{58} &
\end{tabular}
\caption{Type II error rates in PARCS compared to CUSUM$_\text{ML}$ for different lengths of the time series; underline, method with higher detection rate; nominal $\alpha$ level, 0.05 for both methods.} \label{tab:01}
\end{table}

\subsection{Detecting Multiple CPs in Univariate Data} \label{sec:0302}

For the scenario with multiple CPs, we assess the performance of PARCS in comparison to the CUSUM method with standard binary segmentation \citep{Bai1997, Scott1974} for univariate data with white Gaussian noise. Standard binary segmentation is known to mislocate CPs in some scenarios, but modifications to the segmentation procedure have been proposed for solving this problem \citep{Fryzlewicz2014}. We show that PARCS provides an alternative approach. We then discuss a fundamental practical problem in statistical testing when using segmentation methods in general that is avoided by PARCS. Through comparison with standard binary segmentation, we illustrate conditions under which using such methods becomes infeasible. We then consider temporally dependent noise in univariate time series.

The \emph{binary segmentation} method \citep{Bai1997, Scott1974} for detecting multiple CPs proceeds as follows \citep[pseudocode can be found in][]{Fryzlewicz2014}: If, according to a CUSUM test criterion, a CP $\hat{c}_1$ is detected over the full time series, the series is partitioned at $\hat{c}_1$. The procedure is repeated on the resulting left and right segments, potentially returning two additional CPs, $\hat{c}_{21}$ and $\hat{c}_{22}$, respectively. The procedure is terminated when no more CPs are detected after subsequent partitioning. Similar to the single CP case, we use bootstrap testing in deciding the significance of a CP at each stage. In the present context, we will refer to this CUSUM-based binary segmentation method simply by `CUSUM'.

\begin{figure}[!t]
\centering
\includegraphics[width=\linewidth]{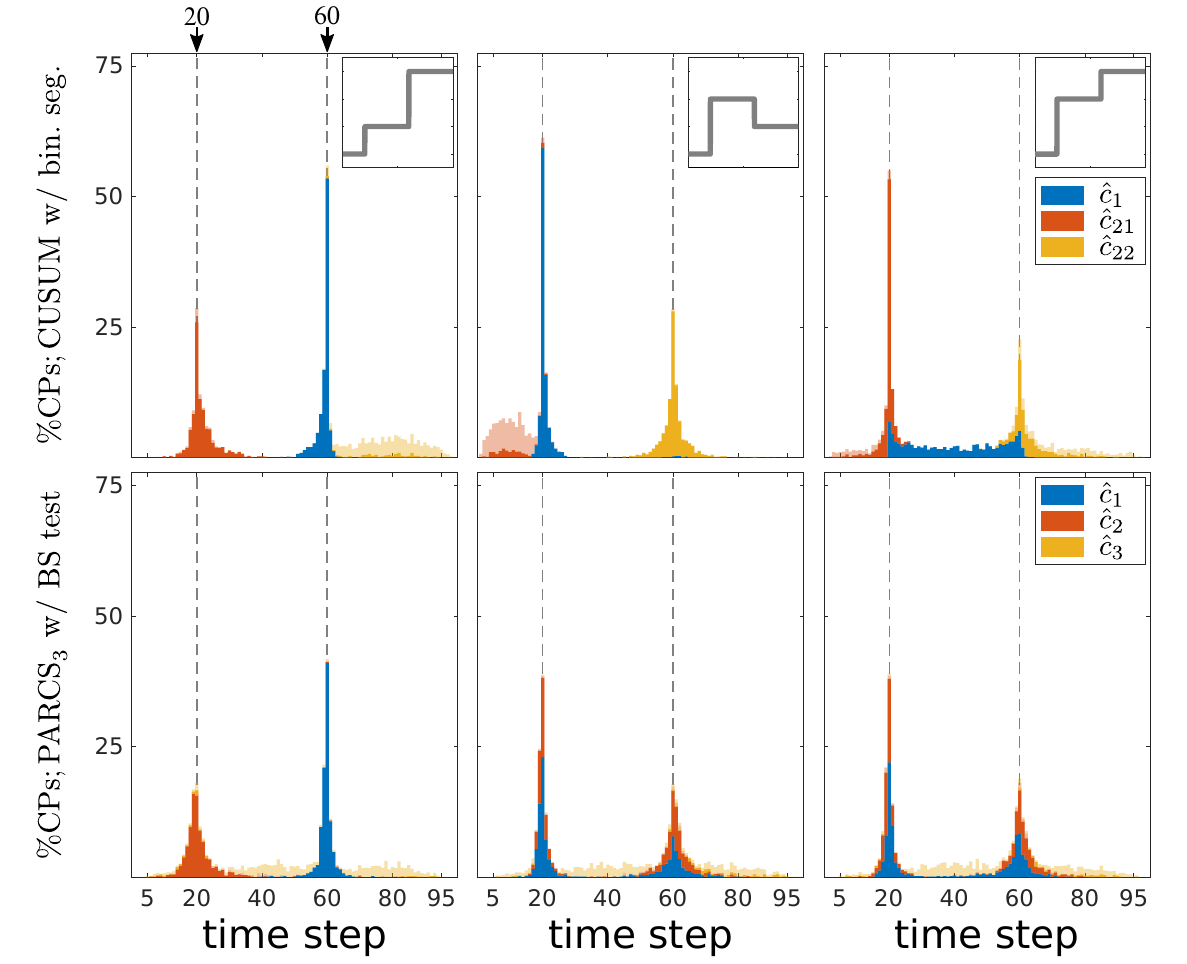}
\caption{Comparing PARCS to CUSUM with binary segmentation for multiple CP detection; stacked histograms of correct detection rates for CUSUM's $\hat{c}_1,\hat{c}_{21},\hat{c}_{22}$ (top) and PARCS' $\hat{c}_1,\hat{c}_2,\hat{c}_3$ (bottom) over 1000 realisations; transparent bars show candidate CPs excluded by the permutation test; dashed grey, ground truth CPs; top inset, deterministic component of time series for the respective column's scenario; left, centre, and right panels refer to first, second, and third scenario, respectively.}
\label{fig:05}
\end{figure}

Three different processes with white Gaussian noise, $\sigma = 1.0$, of the form given by Eq.~\ref{eq:xM} and of length $T = 100$ are simulated for 1000 realisations each. A baseline $b = 0$ and two CPs at time steps $c_1 = 20$ and $c_2 = 60$ are set in all three scenarios. Weights $(w_1,w_2)$ are set to $(1,2)$, $(2,-1)$ and $(2,1)$ (see insets in top row of Figure \ref{fig:05}). For the present comparison, binary segmentation is terminated after at most one partitioning, as this completely suffices to compare the methods (note that this allows CUSUM to detect up to three potential CPs, with only two present in the series). Similarly, the PARCS$_3$ model is estimated, and both methods use the corresponding permutation bootstrap test with $B = 10000$ and $k = 1$. In order to compare type I and type II error rates between the two methods, we set nominal $\alpha$ levels to 0.05 and 0.30 for CUSUM and PARCS, respectively, which is expected to return about the same factual type I error rates of 5\% for both methods in series of length $T \in [26, 100]$, according to Figures \ref{fig:02}E-G.

A first look at Figure \ref{fig:05} suggests that both CUSUM and PARCS detect CPs that are close to the ground truth. A more detailed comparison with respect to type I and type II error rates and the quality of CP detections is provided in Table \ref{tab:02}. The quality of detections using \emph{accuracy scores} is defined as the correct detection rate within a $\pm 5\%T$ range from the ground truth CP location, adjusted for type I errors by an additive term of $-\hat{\alpha}/M$, where $\hat{\alpha}$ is the factual (empirical) $\alpha$ level. This way, the accuracy score is an overall performance measure that takes into account both type I and type II errors, and how far off the detected CP is from the true one.

The first scenario (left panels in Figure \ref{fig:05}) has the hardest parameter setting, since $c_1$ is both more peripheral and smaller in magnitude than $c_2$. CUSUM first detects the easier CP, followed by detecting $c_1$ at the left hand segment as $\hat{c}_{21}$, but with a lower accuracy score as confirmed in Table \ref{tab:02}. Similarly, PARCS returns $c_2$ and $c_1$ as the first and second rank CPs, respectively, with accuracy scores higher than those of CUSUM. The relatively low accuracy scores at detecting $c_1$ in both methods are due to its peripheral location and small magnitude. Both type I and type II error rates are markedly lower in PARCS.

\begin{table}[t!]
\centering
\begin{tabular}{c | c | c c | c c}
$T$ & \textbf{method} & \multicolumn{2}{c|}{\textbf{error rate}} &
\multicolumn{2}{c}{\textbf{accuracy score}} \\
& & type I & type II & $c = 20\%T$ & $c = 60\%T$ \\
\hline \hline
& CUSUM & 10 / 14 / 41 & 08 / 02 / 01 & 72 / 92 / 77 & 91 / \underline{78} / 45 \\
100 & & & & & \\
& PARCS & \underline{02} / \underline{03} / \underline{02} & \underline{04} / \underline{00} / \underline{01} & \underline{80} / \underline{96} / \underline{95} & \underline{96} / 74 / \underline{76} \\
\hline
& CUSUM & 12 / 18 / 19 & 27 / 21 / 11 & 34 / 56 / 61 & 74 / 38 / 16 \\
50 & & & & & \\
& PARCS & \underline{04} / \underline{04} / \underline{03} & \underline{13} / \underline{02} / \underline{05} & \underline{51} / \underline{82} / \underline{82} & \underline{85} / \underline{52} / \underline{52} \\
\hline
& CUSUM & 12 / 22 / 18 & 41 / 44 / 16 & 28 / 44 / 71 & 77 / 26 / 25 \\
26 & & & & & \\
& PARCS & \underline{06} / \underline{07} / \underline{06} & \underline{24} / \underline{09} / \underline{10} & \underline{37} / \underline{75} / \underline{76} & \underline{79} / \underline{47} / \underline{47}
\end{tabular}
\caption{Comparing PARCS to CUSUM with binary segmentation for multiple CP detection for different lengths of the time series; error rates and accuracy scores are rounded; triplet, scenarios 1 / 2 / 3; underline, method with lower error rate or higher accuracy score; nominal $\alpha$ levels, 0.05 and 0.30 for CUSUM and PARCS, respectively.} \label{tab:02}
\end{table}

The second and third scenarios (centre and right panels in Figure \ref{fig:05}, respectively) are easier in terms of ground truth parameter settings, since the more peripheral CP $c_1$ has the larger magnitude. In the second scenario (centre panels in Figure \ref{fig:05}), the two methods are comparable with regards to their overall accuracy scores as defined above (many detections lie outside the $\pm 5\%T$ accuracy score range, especially for $c_2$), but PARCS has the lower type I and type II error rates. CUSUM has a higher rate of false discoveries than in the first scenario. Its first detection is the higher magnitude CP, which comes with a lower accuracy than PARCS, followed by a detection at the right hand segment with a higher accuracy than PARCS. The relatively low accuracy rates for detecting $c_2$ in both methods are due to its small magnitude. The third scenario (right panels in Figure \ref{fig:05}) is an example of a setting in which standard binary segmentation may fail in correctly allocating CPs \citep[see][for a binary segmentation approach that solves this problem]{Fryzlewicz2014}. While the performance of PARCS remains about the same as in the second scenario, CUSUM's first detection diverges from either of the two ground truth CPs in a large number of realisations (see top-right panel in Figure \ref{fig:05}). The large type I error rate (more than three times that in the first and second scenarios) markedly reduces the accuracy scores, which are substantially lower than for PARCS for both ground truth CPs (see Table \ref{tab:02}). Another factor behind the high type I error rates of any iterative procedure including binary segmentation is that the same CP could be detected again at later iterations when an earlier detection is slightly biased.

We now compare the impact of shorter series on the performance of binary segmentation methods and PARCS. We consider 1000 noise realisations from each of the three scenarios with $T \in \{ 50, 26\}$. Two ground truth CPs are set to the same relative location within the time series as for $T = 100$. As seen in Table \ref{tab:02}, both type I and type II error rates increase in both methods with the decrease in series length, with the exception of type I error rates for CUSUM in the third scenario. PARCS is consistently the superior method, having both higher statistical power and less false discoveries than CUSUM. While accuracy scores predictably decrease with shorter series length, comparison between the two methods remains qualitatively similar to the $T = 100$ case in the first and third scenario, while there is a marked change in the second scenario: While CUSUM is slightly superior in accurately detecting $c_2$ for $T= 100$, PARCS progressively surpasses CUSUM in accuracy as series length decreases. This behaviour is a result of deterioration in the power of the bootstrap test statistic for shorter series. Not only does the overall sample size decrease, but CUSUM in the second scenario with $T = 26$ is tasked after a potential first detection of $c_1$ with bootstrapping the CUSUM test statistic on segments as short as 4 or 5 time steps only, which is statistically infeasible, either when using bootstraps or approximate parametric tests \citep{Cho2015, Fryzlewicz2014, Olshen2004}. We stress that this is a fundamental drawback to any method that relies on partitioning, and is not specific to standard binary segmentation.

The limitations of binary segmentation methods become more obvious when noise is temporally dependent. For instance, given a time series of length $T = 100$ with parameters as in the second scenario and an order-2 MA noise process, blocks of size $ k \simeq 3$ are required for proper block-permutation. If CUSUM first detected $c_1 = 20$ accurately, $\hat{c}_1 = 20$, the left hand segment would be only 20 time steps long. This allows for only 7 blocks, yielding 5040 possible permutations. This number drops to 720 permutations had $\hat{c}_1$ been detected only 2 time steps further to the left, which makes it hard to approximate the EDF of the CUSUM statistic reliably. In addition, specifying the block size first requires estimating the MA process order by approximating the $H_0$-conform time series (see Algorithm \ref{alg:02}), but potential CPs are not known \emph{a priori}, due to the recursive nature of binary segmentation methods.

\begin{figure}[!t]
\centering
\includegraphics[width=\linewidth]{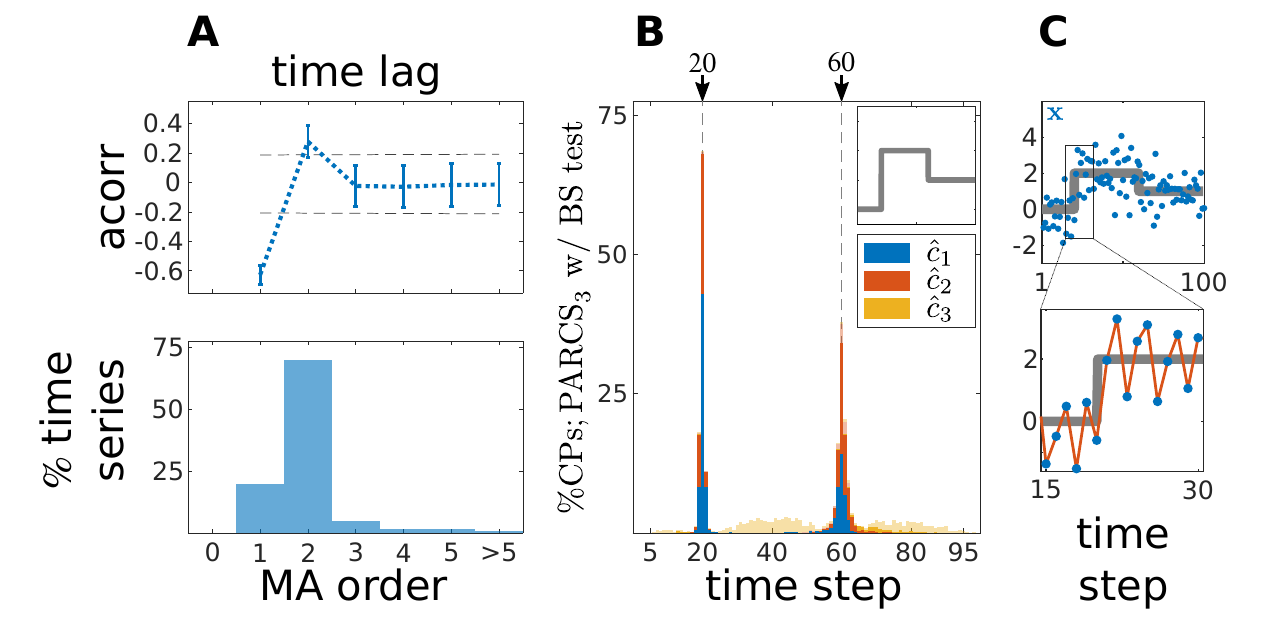}
\caption{Multiple CP detection in temporally dependent data from the second scenario; (\textbf{A}) estimating MA order; (top) average autocorrelation over time series realisations for different time lags $\pm\;\mathrm{s.d.}$; dashed grey, 95\% confidence interval; (bottom) ratio of 1000 realisations with a given estimated order; (\textbf{B}) stacked histograms of correct detection rates over 1000 realisations; transparent bars show candidate CPs excluded by the permutation test; dashed grey, ground truth CPs; inset, deterministic component of time series; (\textbf{C}) deterministic component of the time series (grey) superimposed on an exemplary time series (blue); bottom panel shows a close up over 16 data points around $c_1$; red line highlights oscillation due to the negative MA coefficient.}
\label{fig:06}
\end{figure}

Given these considerations, we focus on PARCS only as we now move over to the case of detecting multiple CPs in series with temporally dependent noise. 1000 noise realisations of length $T=100$ are drawn from an order-2 MA process with $\sigma = 0.7$ and coefficients $\kappa_1 = -0.5 / \sigma$ and $\kappa_2 = 0.4 / \sigma$. Other parameters are as in the previous analysis. In Figure \ref{fig:06}, we illustrate distributions of correct detections for the second scenario only (with the other two scenarios qualitatively comparable to their counterparts in Figure \ref{fig:05}), but error and accuracy rates on these scenarios are reported as well. After removing the three step changes following PARCS$_3$ CP detection, the majority of residual time series (70\% as shown in bottom panel in Figure \ref{fig:06}A) had an autocorrelation that cuts off at the correct order of the ground truth MA$(2)$ noise process, i.e. with $\mathrm{acorr}(\mathbf{x}_0 ; 2)$ being the last coefficient that lies outside the 95\% confidence bounds (dashed lines; top panel in Figure \ref{fig:06}A; results for the first and third scenarios are comparable). Block-permutation bootstrap testing with nominal $\alpha = 0.05$ and $B = 10000$ is carried out on these series with blocks of size 3 and, for other time series, according to the estimated order in Figure \ref{fig:06}A (with an upper bound of 10 on block size). Exactly two CPs are detected in more than 99.5\% of realisations in all scenarios. Figure \ref{fig:06}B shows that the distribution of correct detections for the second scenario is largely concentrated around the ground truth CPs. Accuracy scores in each scenario are, respectively, 96\%, 99\%, and 99\% for $c_1$ and 99\%, 89\%, and 89\% for $c_2$. Note also the oscillation in the example realisation in Figure \ref{fig:06}C, which results from dependent noise with a negative MA coefficient $\kappa_1$.

\subsection{Detecting Multiple CPs in Multivariate Data} \label{sec:0303}

\begin{figure}[!t]
\centering
\includegraphics[width=\linewidth]{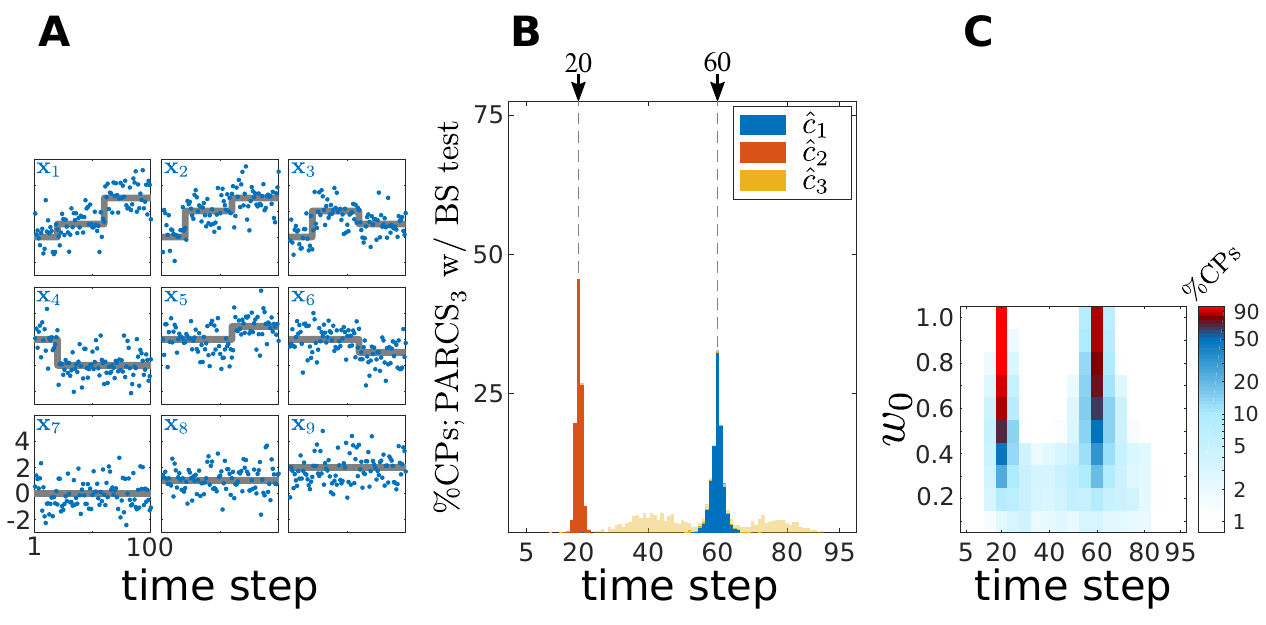}
\caption{Multiple CP detection in spatially independent, multivariate data; (\textbf{A}) deterministic component of the different covariates in the time series (grey) superimposed on an exemplary time series (blue); (\textbf{B}) stacked histograms of correct detection rates over 1000 realisations; transparent bars show candidate CPs excluded by the permutation test; dashed grey, ground truth CPs; (\textbf{C}) stacked histograms of correct detection rates over 1000 realisations given different signal-to-noise ratios (controlled by $w_0$), and binned with 5 time step windows; rates are logarithmically scaled as indicated by the colour bar.}
\label{fig:07}
\end{figure}

The PARCS method's ability to detect multiple CPs in spatially independent, multivariate time series is demonstrated in Figure \ref{fig:07} on 1000 realisations of length $T=100$ with $N=9$ covariates and white Gaussian noise, $\sigma = 1$. Parameters are set to $\mathbf{b} = (0, 0, 0, 2, 2, 2, 0, 1, 2)$, $c_1 = 20$, $\mathbf{w}_1 = w_0 \cdot (1, 2, 2,-2, 0, 0, 0, 0, 0)$, $c_2 = 60$ and $\mathbf{w}_2 = w_0 \cdot (2, 1,-1, 0, 1,-1, 0, 0, 0)$. The scaling parameter $w_0$ controls signal-to-noise ratio in the time series and is initially set to 1.0. Given these parameter values, the two CPs are not represented in all covariates of the time series, as exemplified in Figure \ref{fig:07}A, rendering CPs harder to detect from the averaged univariate time series (with steps differing in sign across the covariates partially cancelling each other, resulting in small weights, $\langle \mathbf{w}_1 \rangle = 3/9$ and $\langle \mathbf{w}_2 \rangle = 2/9$, for the resulting univariate time series).

Following PARCS$_3$ CP detection, augmented by a bootstrap test with nominal $\alpha = 0.05$, $B = 10000$ and $k = 1$, exactly two CPs are detected in 99.9\% of realisations. Accuracy scores are 99.8\% and 98\% for $c_1$ and $c_2$, respectively. The lower variance in $c_1$ detections, as seen in Figure \ref{fig:07}B, is due to the higher average absolute weight $ \bigl \langle |\mathbf{w}_1 | \bigr \rangle = 7/9$ compared to $\bigl \langle | \mathbf{w}_2 | \bigr \rangle = 6/9$.

We then test the method's performance for smaller signal-to-noise ratios with $w_0 \in \{ 1.0 , 0.9 , \dots , 0.1 \}$. Figure \ref{fig:07}C shows that correct detection rates within a $\pm 2\%T$ range from the ground truth CPs for different values of $w_0$ remain above 50\%, even for magnitudes as small as $w_0 = 0.5$. These rates are a result of PARCS leveraging CP information from multiple covariates simultaneously, rather than depleting the signal through averaging.

\subsection{Detecting Neural Events that Reflect Learning} \label{sec:0304}

A previous study by one of the authors and colleagues exemplifies the practical value of change point detection in neuroscience \citep{Durstewitz2010}. These authors demonstrated that acquiring a new behavioural rule in rats is accompanied by sudden jumps in behavioural performance, which in turn is reflected in the activity of neural units recorded simultaneously in the medial prefrontal cortex (mPFC). In the current section, we revisit part of these data to showcase PARCS in a real data scenario.

Before moving to the demonstration, it is important to note that the data in question are not normally distributed and potentially include linear trends \citep{Durstewitz2010} not accounted for by the step models in Eqs.~\ref{eq:model1}, \ref{eq:xM} and \ref{eq:xMN}. As such, some preprocessing may be necessary for a statistical analysis that is more consistent with the step model assumptions \citep[this may include detrending and potentially some mild smoothing with Gaussian kernels; see][]{Durstewitz2010}. However, to keep the present demonstration simple, PARCS was applied directly to the data with minimal preprocessing, which only involves square-root-transforming the neural count data for bringing them closer to a Gaussian distribution and stabilising the variance \citep{Kihlberg1972}.

\begin{figure}[!t]
\centering
\includegraphics[width=\linewidth]{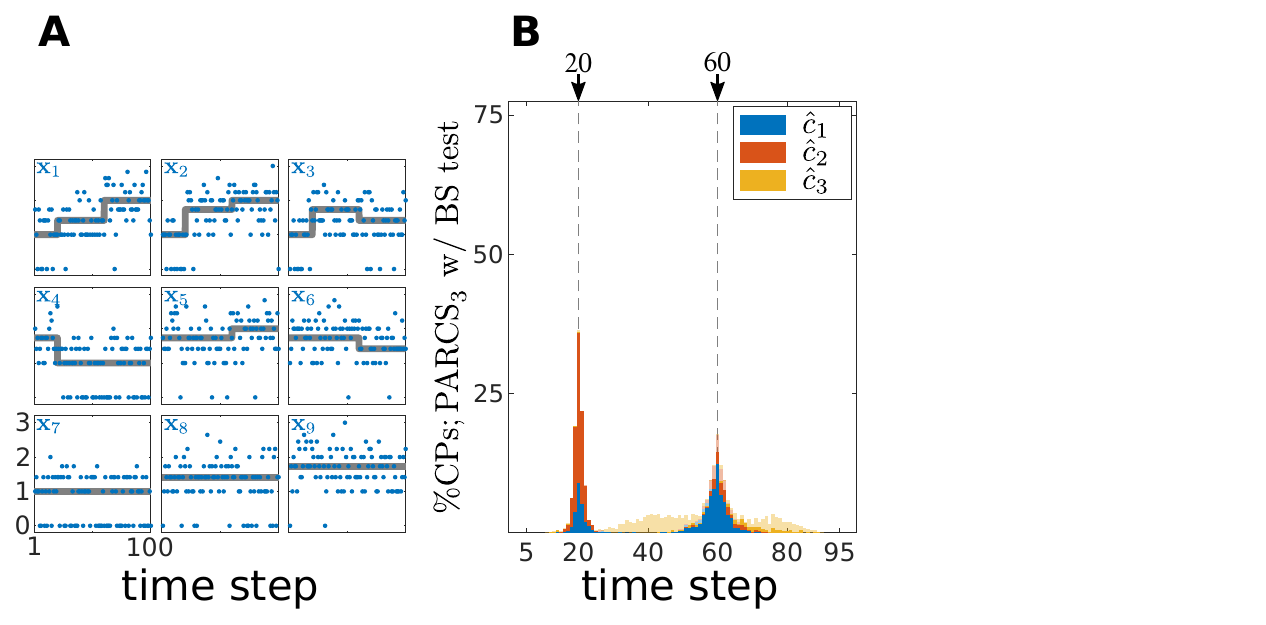}
\caption{Multiple CP detection in spatially independent, multivariate, Poisson data; (\textbf{A}) deterministic component of the different covariates in the time series (grey) superimposed on an exemplary time series (blue); $y$-axis, square-root-transformed spike counts; (\textbf{B}) stacked histograms of correct detection rates over 1000 realisations; transparent bars show candidate CPs excluded by the permutation test; dashed grey, ground truth CPs.}
\label{fig:08}
\end{figure}

In order to show that PARCS can still return reasonable CP estimates under these non-Gaussian conditions, we first test its performance on simulated spike count data, before applying it to the empirical data. We simulate 1000 realisations of length $T=100$ with $N=9$ covariates according to a Poisson process. Parameters are set to $\mathbf{b} = (1, 1, 1, 3, 3, 3, 1, 2, 1)$, $c_1 = 20$, $\mathbf{w}_1 = (1, 2, 2,-2, 0, 0, 0, 0, 0)$, $c_2 = 60$ and $\mathbf{w}_2 = (2, 1,-1, 0, 1,-1, 0, 0, 0)$. This choice of parameters results in average firing rates that are comparable in their means to the white Gaussian noise case (cf.\ Figures \ref{fig:07}A and \ref{fig:08}A) and to the low firing rates often observed in mPFC neurons. One obvious diversion from Gaussian assumptions in the case of Poisson noise is that the variance is not constant anymore, but is equal to the means within each of the segments separated by true CPs. Following square-root-transforming the data and PARCS$_3$ CP detection, augmented by a bootstrap test with nominal $\alpha = 0.05$, $B = 10000$ and $k = 1$, exactly two CPs are detected in 92\% of realisations. Accuracy scores are 98\% and 70\% for $c_1$ and $c_2$, respectively. The lower accuracy scores compared to the white Gaussian noise scenario are due to the lower signal-to-noise ratios, resulting from the increase in noise variance with firing rates (cf.\ Figures \ref{fig:07}B and \ref{fig:08}B). Nevertheless, these results sufficiently justify the use of PARCS in the present context￹.

\begin{figure}[p]
\centering
\includegraphics[width=\linewidth]{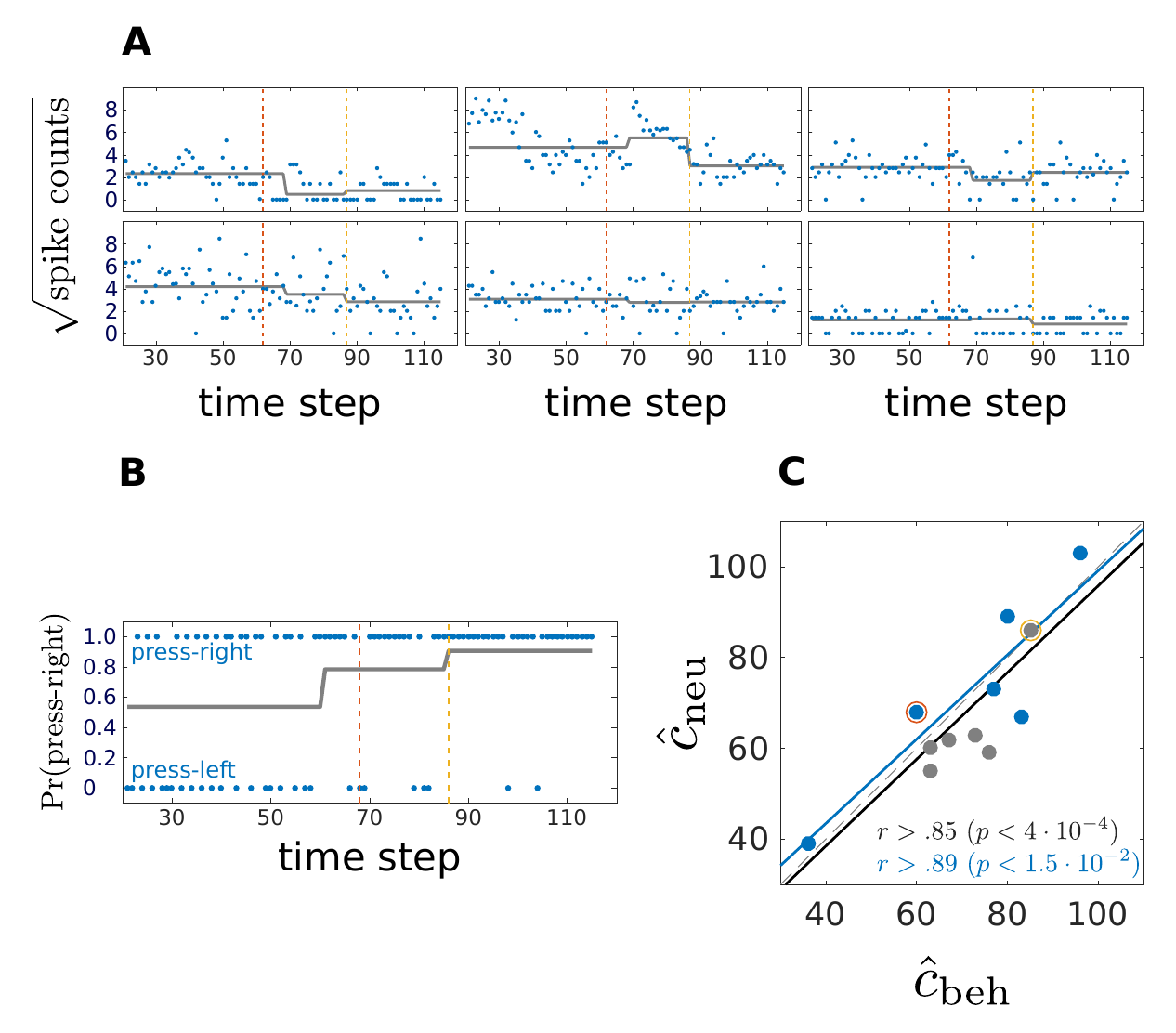}
\caption{Comparing behavioural and mPFC neural CPs; (\textbf{A}) blue, square-root-transformed spike count data in the three seconds following cue onset from 6 representative mPFC units of one rat; grey, mean as estimated by inverting the neural multiple response PARCS$_2$ model. Note potential CP in top-centre unit which was not detected by PARCS$_2$ since it did not contribute strongly to population-wide CPs; dashed lines, behavioural CPs from the same animal; (\textbf{B}) blue, lever press at each trial; this animal is rewarded for pressing the right lever during the spatial rule; grey, probability of pressing right lever as estimated by inverting the behavioural PARCS$_2$ model; dashed lines, neural CPs from the same animal (see A); (\textbf{C}) relating behavioural and neural CPs; blue, behavioural CPs with higher weight; $r$, correlation coefficients as computed over all 12 data points (black) and over those where behavioural CPs have the higher weight (blue); $p$-values, significance levels of corresponding $r$; black and blue lines, respective least-square linear regression fits to the two sets of data points; red and yellow circles, neural and behavioural CP pairs from the exemplary animal in A and B, respectively.}
\label{fig:09}
\end{figure}

We now turn to the experimentally obtained dataset. Six animals were trained on a two-choice deterministic operant rule switching task which proceeds as follows: At the beginning of the session, the animal follows a previously acquired behavioural rule whereby it responds to a visual cue by a lever press for attaining a reward (visual rule). Unknown to the animal, reward contingencies are switched after 20 trials to a novel spatial rule, in which attaining the reward requires pressing a certain baited lever (right or left), regardless of the visual cue. The session is terminated when the animal reaches a preset criterion that indicates that the new rule behaviour has been learnt. In addition to the binary behavioural data of lever presses over trials, spike counts emitted by mPFC units during the 3 seconds following cue onset were collected through single unit recording techniques. Neural and behavioural data from one animal are shown in Figure \ref{fig:09}A and \ref{fig:09}B, respectively. Trials corresponding to the steady state visual and spatial rule (first and last 20 trials, respectively) are not considered in the analysis.

Time series with one or two significant CPs were described in the original study by\cite{Durstewitz2010}, so PARCS$_2$ models are estimated for each animal for both the multivariate neural data (multiple response PARCS model; Figure \ref{fig:09}A) and the univariate behavioural data (Figure \ref{fig:09}B), in addition to PARCS$_1$ models for the behavioural data as summarised in Figure \ref{fig:09}C. As shown in Figure \ref{fig:09}B, one neural CP in that exemplary animal matches its behavioural counterpart. The second neural CP, while not as close to its behavioural counterpart, is only 7 trials apart, and the two are highly correlated across animals, as shown in Figure \ref{fig:09}C, concurring with the original findings of \cite{Durstewitz2010}. Besides the significant correlation, the corresponding black linear regression line lies very close to the diagonal, indicating that neural and behavioural CPs are not only correlated, but are almost equal. Moreover, those authors report that data from many animals contain at most a single CP (also note the low weight of one of the CPs as estimated from one of the animals using PARCS$_2$). Comparing the PARCS$_1$ behavioural CP to its neural counterpart (blue circles in Figure \ref{fig:09}C) shows that correlation remains high and significant. A sample size of $s>5$ is usually recommended for evaluating the significance of a correlation. However, the corresponding linear regression line is also close to the diagonal, in further support for the reliability of this result, and in agreement with the original results despite different procedures: For the neural data in the original study, CUSUM-based detection was performed on a multivariate discrimination statistic defined across the whole neural population, while here, the model was determined directly from the multiple spike count data.

\section{Discussion} \label{sec:04}

In the current article, we introduced PARCS, a method for detecting multiple step changes, or CPs, in potentially multivariate, temporally dependent data, supported by a bootstrap-based nonparametric test. We also showed that PARCS substantially reduces centre bias in estimating CPs compared to the most basic specification of the CUSUM method, and presented conditions under which it compares to or outperforms the maximum-likelihood CUSUM statistic. Furthermore, we demonstrated that PARCS may achieve higher sensitivity (statistical power) than CUSUM-based methods while at the same time having lower type I errors in multiple CP scenarios, mainly because PARCS can make use of the full time series while CUSUM-based methods rely on segmenting the time series for detecting multiple CPs. We finally confirmed previous results pertaining to the acquisition of a new behavioural rule and the role of the medial prefrontal cortex in this process.

As already apparent from some of our simulation studies, the basic PARCS method as introduced here leaves room for improvement. In the presence of a single CP, we showed that PARCS strongly reduces the amount of bias toward the centre that results from the direct application of the most basic form of the CUSUM locator statistic. Theoretically-grounded modifications to the CUSUM transformation that reduce this amount of bias rely on down-weighing more centrally-located points \citep{Kirch2007}. As shown with PARCS, this problem is not quite as severe. Nevertheless, since PARCS approximates the CUSUM transformation using a regression model, similar down-weighing could be incorporated into the PARCS procedure as well by using \emph{weighted least squares} instead of regular least squares \citep{Hastie2009}, which is a straightforward amendment. 
Furthermore, the PARCS method currently requires a liberal guess of the number $M$ of CPs in advance, followed by refinements through nonparametric bootstrap testing. It is desirable, however, especially when no prior information on $M$ is available, to have statistical tests as termination criteria for the forward and backward stages. In adaptive regression spline methods \citep{Friedman1991, Friedman1989, Stone1997}, there is strong empirical evidence \citep{Hinkley1969, Hinkley1971b} backed by theoretical results \citep{Feder1975} that the difference in residual mean-square-error between two nested models that differ in one additional knot is well approximated, albeit conservatively, by a scaled $\chi^2$ statistic on 4 degrees of freedom \citep{Friedman1991}. This led to one nonparametric termination recipe that is based on generalised cross-validation \citep{Craven1978}. Another approach is to infer the piecewise linear regression model with the aid of a parametric test for specifying the number and location of knots, without recourse to iterative procedures \citep{Liu1997}. Unfortunately, neither approach is directly applicable to PARCS, since they both require assumptions that are not met in the CUSUM-transformed time series. The CUSUM transformation of the time series is a nonstationary ARMA$(1,q)$ process. Deriving reasonable generalised cross-validation \citep{Craven1978, Friedman1991, Friedman1989}, F-ratio \citep{Durstewitz2017, Hastie2009} or parametric \citep{Liu1997} test statistics require currently unknown corrections to those tests which account for nonstationarity and the particular form of the ARMA model underlying the CUSUM-transformed data.

When multiple CPs are present in the data, PARCS can outperform standard binary segmentation \citep{Bai1997, Scott1974}. Other segmentation methods also solve the problem of mislocating CPs inherent in the standard procedure \citep{Cho2015, Fryzlewicz2014, Olshen2004}. Wild binary segmentation \citep[WBS;][]{Fryzlewicz2014}, for instance, relies on sampling local CUSUM transformations of randomly chosen segments of the time series. The candidate CP with the largest value among sampled CUSUM curves is returned to be tested against a criterion, followed by binary segmentation. WBS is preferable to PARCS in that its test statistic and termination criterion when noise is independent are backed up by rigorous theory, and may be the favourable method when segments are large enough for the test statistic to converge. If series of only limited length are available, however, WBS may run into similar problems as standard binary segmentation for CUSUM, since each detection is still followed by partitioning the data further. WBS also, to the best of our knowledge, currently lacks a thorough analysis on the behaviour of its test statistic for dependent data. It is tempting to speculate on the potential for a hybrid method that capitalises on the desirable features of both methods. Computational demands arise in WBS from the need to choose segment range parameters by sampling few thousand CUSUM curves to which PARCS may offer an easy and efficient workaround: \cite{Fryzlewicz2014} demonstrated that the optimal WBS segment choice is the segment bounded by the two CPs closest to the target CP from each side. PARCS could thus provide an informed selection of boundaries by returning candidate CPs in the data and use these to demarcate segments, rather than random sampling as in WBS.

In dealing with multivariate data, recent methods tackled the computational demands of having a large number of covariates and sparse CP representations \citep{Cho2015, Wang2018}. These methods rely on low dimensional projections of the multivariate CUSUM curve that preserve the CPs and follow this projection by a binary segmentation method. Since PARCS for multivariate time series is also based on the CUSUM transformation, it is straightforward to leverage the computational savings provided by such projection methods in reducing the dimensionality of the PARCS input, while avoiding the drawbacks of binary segmentation methods. This may offer a route for extending PARCS to the important case of multivariate CP detection in mutually dependent time series with spatial dependence, a configuration which these projection methods also consider \citep{Cho2015, Wang2018}. Alternatively, nondiagonal covariance structure in multivariate series may be accounted for by extending the PARCS formulation to the multivariate regression spline realm \citep{Friedman1991, Stone1997}.

Finally, when analysing the neural and behavioural data during the rule switching task, we mentioned that data may also contain trends that are not accounted for by step change time series models \citep{Durstewitz2010}. Caution must be made when analysing real data using CP detection methods in that these methods, PARCS included, assume a step change model underlying the generation of the data and hence may attempt to approximate trends and other nonstationary features by a series of step changes, a point made more explicit by \cite{Fryzlewicz2014} \citep[][therefore removed trends around candidate CPs first]{Durstewitz2010}. Hence, to avoid wrong conclusions with respect to the source and type of nonstationarity in experimental time series, it may be necessary to either augment change point detection by adequate preprocessing \citep{Durstewitz2010} or to generalise time series models for CP detection to include other forms of nonstationarity.

\section*{Author Contributions}

HT, DD conceived study; HT developed and implemented methods; HT carried out simulations; HT analysed data; HT, DD interpreted results; HT prepared figures; HT wrote manuscript; HT, DD revised and finalised manuscript.

\section*{Funding}

This research was funded by grants to DD by the German Research Foundation (DFG) (SPP1665, DU 354/8-2) and through the German Ministry for Education and Research (BMBF) via the e:Med framework (01ZX1311A \& 01ZX1314E).

\section*{Acknowledgements}

The authors thank Dr.\ Georgia Koppe and Dr.\ Eleonora Russo for discussions and Dr.\ Jeremy Seamans for providing the neural and behavioural data.

\section*{Data Availability Statement}

Method implementation will be freely available on an online repository upon publication.

\bibliographystyle{apacite}

\end{document}